\documentclass{emulateapj}
\usepackage{apjfonts}
\usepackage{color}

\shorttitle{$K_S$ and IRAC Selected EROs}
\shortauthors{Wang, Barger, \& Cowie}

\begin{document}

\title{A $K_S$ and IRAC Selection of High-Redshift Extremely Red Objects\footnotemark[1]}

\author{Wei-Hao Wang\altaffilmark{2}, 
Amy J.\ Barger\altaffilmark{3,4,5},
and Lennox L.\ Cowie\altaffilmark{5}}

\footnotetext[1]{Based on observations obtained at the Canada-France-Hawaii Telescope (CFHT), 
which is operated by the National Research Council of Canada, the Institut National des Sciences 
de l'Univers of the Centre National de la Recherche Scientifique of France, and the University of Hawaii.}
\altaffiltext{2}{Academia Sinica Institute of Astronomy and Astrophysics, 
P.O. Box 23-141, Taipei 10617, Taiwan}
\altaffiltext{3}{Department of Astronomy, University of Wisconsin-Madison, 
475 North Charter Street, Madison, WI 53706}
\altaffiltext{4}{Department of Physics and Astronomy, 
University of Hawaii, 2505 Correa Road, Honolulu, H  96822}
\altaffiltext{5}{Institute for Astronomy, University of Hawaii, 
2680 Woodlawn Drive, Honolulu, HI 96822}

\begin{abstract}
In order to find the most extreme dust-hidden high-redshift galaxies, 
we select 196 extremely red objects in the $K_S$ and IRAC bands (KIEROs, $[K_{s}-4.5\mu \rm m]_{\rm AB}>1.6$) 
in the 0.06 deg$^2$ GOODS-N region.  This selection avoids the Balmer breaks of galactic spectra at $z<4$
and picks up red galaxies with strong dust extinction.  The photometric redshifts of KIEROs are between
1.5 and 5, with $\sim70\%$ at $z\sim2$--4.  KIEROs are very massive, 
with $M_\star\sim10^{10}$--$10^{12} M_\sun$.  They are optically faint and usually cannot
be picked out by the Lyman break selection.  On the other hand, the KIERO 
selection includes approximately half of the known millimeter and submillimeter galaxies in 
the GOODS-N.  Stacking analyses in the radio, millimeter, and submillimeter all show that
KIEROs are much more luminous than average 4.5~$\mu$m selected galaxies.  
Interestingly, the stacked fluxes for ACS-undetected KIEROs in these wavebands are 2.5--5 times 
larger than those for ACS-detected KIEROs.  With the stacked radio fluxes and the local radio--FIR 
correlation, we derive mean infrared luminosities of 2--$7\times10^{12}L_\sun$ and mean star formation 
rates of 300-1200 $M_\sun$ yr$^{-1}$ for KIEROs with redshifts.  We do not find evidence of a 
significant subpopulation of passive KIEROs. 
The large stellar masses and star formation rates 
imply that KIEROs are $z>2$ massive galaxies in rapid formation. Our results show that a large sample
of dusty ultraluminous sources can be selected in this way and that a large fraction of
high-redshift star formation is hidden by dust.
\end{abstract}

\section{Introduction}
Deep imaging in the submillimeter opened a new window for studying 
high-redshift galaxies \citep{smail97,hughes98,barger98}.  The brighter submillimeter
galaxies (SMGs, typical total infrared luminosity greater than $10^{13} L_\sun$) that have radio 
counterparts were found to be at redshifts mostly between 2 and 3 
\citep{chapman03,chapman05}.  Despite the extremely large star formation rates 
(SFRs), typically $>1000$ $M_\sun$ yr$^{-1}$, such galaxies do not emit strongly 
in the rest-frame ultraviolet (UV) and therefore are generally not Lyman break galaxies 
\citep[LBGs; ][]{peacock00,chapman00,webb03}.  (A counter example is a $z=4.55$ 
SMG in \citealp{capak08}, which was originally selected as 
an LBG.)  This implies a large fraction of dusty star 
formation at high redshift being missed by rest-frame UV surveys.  

This raises two general questions:  will we start to see galaxies similar to optically 
selected ones if we are able to probe deeper in the submillimeter, and exactly 
how much high-redshift star formation is missed by optical surveys?  
We are not close to being able to answer either question.  Deep submillimeter 
surveys in lensing clusters are able to
probe dusty galaxies with IR luminosities $\lesssim10^{12} L_\sun$ and
constrain their number density \citep{blain99,cowie02,knudsen08,chen11}, but
the identification of their optical counterparts is not yet complete, and the
sample sizes are very small.  Furthermore, a $z\gtrsim4$ SMG, 
GOODS 850-5, was recently found to be extremely faint in the optical and
near-infrared (NIR) at $\lambda_{\rm obs} < 2.2$~$\mu$m \citep{wang07,wang09}.  If a 
significant fraction of the $z>2$ SMGs in the IR luminosity range
$10^{12}$--$10^{13}$ $L_\sun$ are like GOODS 850-5, then identifying them in the radio and
optical will be challenging even with next-generation instruments.
We therefore seek a NIR diagnosis of extremely optically faint SMGs.  
A general description of the  \emph{Spitzer}
Infrared Array Camera (IRAC) colors of bright SMGs was recently developed 
by \citet{yun08}.  Here we focus on the reddest 
galaxies in the IRAC and $K_S$ bands.

In this paper we present a new color selection of extremely red
objects (EROs) with $K_S$ and IRAC colors of $K_S-4.5$~$\mu$m $>1.6$. 
We refer to such sources as KIEROs.  This selection is motivated by the fact that at least
half of known SMGs are redder than this color (Section~\ref{sec_data}).  
Among all existing ERO selections, the KIERO selection utilizes the longest wavebands 
that are practically accessible for deep imaging.  Unlike some
other selections for high-redshift red objects, the KIERO selection
does not particularly aim at the 4000~\AA~Balmer  breaks in galactic spectra, 
which only enter the $K_S$ band at $z\sim4$.  Instead, this selection aims at
galaxies at $z>2$ whose extremely red colors are likely caused by large dust extinction
(Section~\ref{sec_redshift}). Also because of this, we choose the 4.5 $\mu$m band
instead of the 3.6 $\mu$m band, to be more sensitive to galaxies whose 
spectral energy distributions (SEDs) are red over a broad wavelength range
(cf. sharp spectral breaks). 

The paper is organized as following.
We describe the data and the KIERO selection in Section~\ref{sec_data}, 
the number counts in Section~\ref{sec_density}, the optical and NIR SEDs, 
redshift distribution, and stellar populations in Section~\ref{sec_redshift}, 
and the radio, millimeter, submillimeter, and X-ray properties in 
Section~\ref{sec_prop}.  We identify active galactic nuclei (AGNs) in KIEROs 
in Section~\ref{sec_agn}. We estimate the SFRs and 
SFR densities (SFRDs) of KIEROs in Section~\ref{sec_sfr}.
We compare KIEROs with other high-redshift galaxy populations in Section~\ref{sec_overlap}.
We discuss the roles of KIEROs in galaxy formation in Section~\ref{sec_discuss} and summarize
our results in Section~\ref{summary}.
Throughout the paper, we assume cosmological parameters of $H_0=71$ km s$^{-1}$ Mpc$^{-1}$, 
$\Omega_M=0.27$, and $\Omega_\Lambda=0.73$.  All magnitudes are in the AB 
system, where an AB magnitude is defined as $\rm AB=8.9-2.5\log(\rm flux in Jy)$.

\section{The KIERO Sample}\label{sec_data}

We use our publicly released $K_S$ and IRAC catalogs \citep[hereafter W10]{wang10} of the
Great Observatories Origins Deep Surveys-North (GOODS-N; \citealp{giavalisco04}) field.
The $K_S$ source catalog of W10 is extracted from an extremely deep Canada-France-Hawaii Telescope
$K_S$ image, which has a $1\sigma$ depth of 0.12 $\mu$Jy in the GOODS-N region.
W10 used the GOODS-N \emph{Spitzer} IRAC images at 3.6--8.0~$\mu$m (M.\ Dickinson et al.\ 2011, 
in preparation) to measure the IRAC photometry of the $K_S$ selected sources using a CLEAN-like method
that takes the $K_S$ catalog and image as priors. W10's primary $K_S$ and IRAC
catalog contains 15018 $K_S$ selected sources that are detected at 3.6 and 4.5~$\mu$m 
in the GOODS-N region.  W10's secondary catalog contains 358 $K_S$-faint IRAC detected sources.

%
%
\begin{figure}[t!]
\epsscale{1.0}
\plotone{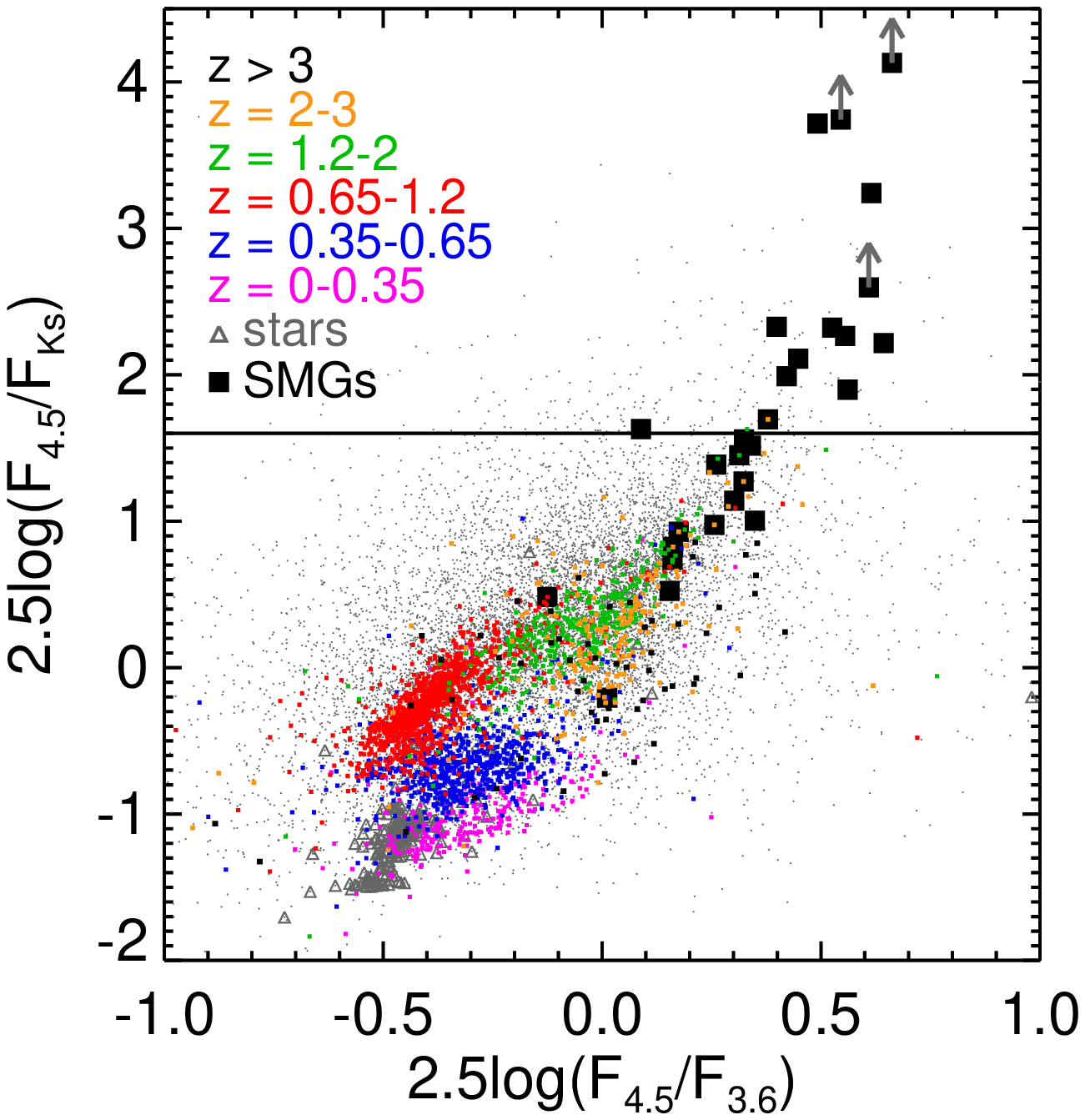}
\caption{$K_S$, 3.6~$\mu$m, and 4.5~$\mu$m color--color diagram.  Small gray dots
are 3 $\sigma$ 4.5~$\mu$m sources in the catalog of W10.  Spectroscopically identified 
sources in \citet{barger08}, including stars, are shown with color symbols.
Radio and submillimeter interferometrically identified 4 $\sigma$ SMGs in the sample of 
Wang et al.\ (2004) and \citet{perera08} are shown with filled squares. The horizontal line 
indicates our selection criterion for KIEROs.
\label{fig_color_color} }
\end{figure}

We show the $K_S$, 3.6~$\mu$m, and 4.5~$\mu$m color--color diagram for sources with $>3\sigma$ 4.5~$\mu$m 
fluxes in W10's primary catalog in Figure~\ref{fig_color_color}.  
To obtain colors of GOODS-N SMGs, we looked for $K_S$ and IRAC counterparts of the 
SMGs in the 4 $\sigma$ SCUBA sample of \citet{wang04} and AzTEC sample of \citet{perera08}.
We only include SMGs with unambiguous radio interferometric identifications 
(Wang et al.\ 2004; \citealp{pope06,chapin09}) or submillimeter interferometric identifications 
(\citealp{iono06,wang07,daddi09a,wang11}; A.\ J.\ Barger et al.\ 2011, in preparation).
There are 28 such SMGs and they are shown with squares in Figure~\ref{fig_color_color}. 
Fourteen of them are redder than $K_S-4.5\mu\rm{m}=1.6$ (horizontal line in the figure)
and three are not even detected in $K_S$ (lower limits in the figure). They occupy a color space 
significantly different than that of most sources detected at both $K_S$ and 4.5~$\mu$m.
Because most spectroscopically identified SMGs are at $z<3$ \citep{chapman03,chapman05}, 
the extremely red colors of these SMGs must be caused by very high redshifts and/or extremely large 
dust extinctions (see Section 4), like in the case of GOODS 850-5 (\citealp{wang07,wang09}).
Such extremely red SMGs are of particular interest.  Since it is known that a variety of 
optical/NIR criteria are required to include all bright SMGs \citep{reddy05},
here we only focus on selecting and understanding the extremely red ones.

In order to find sources similar to the extremely red SMGs but slightly fainter in the submillimeter and therefore
undetected by current submillimeter surveys, we select KIEROs with $K_S-4.5$~$\mu$m $>1.6$.  
We require that the sources be detected at $>3\sigma$ in the 4.5~$\mu$m band and in at least one other IRAC band.
For galaxies detected in the IRAC bands but not in the $K_S$ band (in W10's secondary catalog), we adopt their $1\sigma$
$K_S$ limits.  Although W10's primary catalog is a $K_S$ selected one, the above inclusion of $K_S$-faint
4.5~$\mu$m sources in W10's secondary catalog makes the selection of KIEROs in principle a 4.5~$\mu$m selection.
We visually inspected the IRAC images of all such galaxies
and excluded those whose IRAC fluxes are less reliable, mainly galaxies
affected by very bright stars in the field and galaxies
blended with $\ge 3$ bright nearby galaxies.  These are the limitations of W10's catalogs (see W10 for discussion). 
This reduces the number of selected KIEROs by
$<10\%$ and should not impact our analyses, even if some of them
are truly red sources.  We are left with 104 KIEROs in the main $K_S$-selected
sample and 92 in the $K_S$ undetected sample.   Among the above mentioned 28 identified SMGs in the
GOODS-N, 14 are KIEROs.

\section{Number Counts}\label{sec_density}

\begin{figure}[t!]
\epsscale{1.0}
\plotone{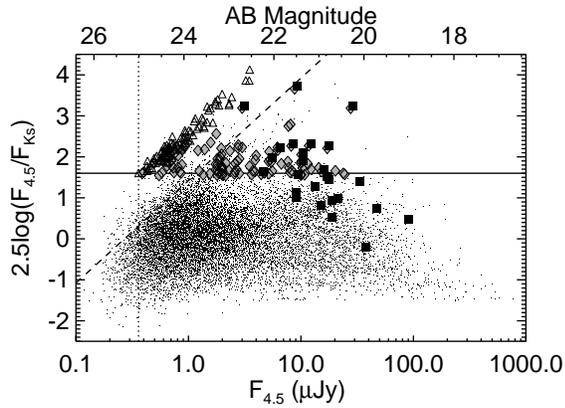}
\caption{$K_S$ and 4.5~$\mu$m color--magnitude diagram.  Dots are field galaxies detected 
(3$\sigma$) in both the $K_S$ and 4.5~$\mu$m bands, as well as at least one other IRAC band.
Diamonds are $K_S$ detected KIEROs, and triangles are $1\sigma$ color lower limits of $K_S$ undetected KIEROs.
Squares are the radio and submillimeter interfeormetrically identified SMGs.
The horizontal solid line shows the selection criterion for KIEROs.
Some of the galaxies above the horizontal line are not designated as KIEROs because their 
4.5~$\mu$m fluxes in W10's catalogs are less reliable, mostly bad photometry affected by very 
bright stars in the field or galaxies blended with $\ge 3$ bright nearby galaxies.
The diagonal dashed line shows the $K_S$ 30\%
completeness limit determined by W10.  The vertical dotted line shows the $3\sigma$
limit of the 4.5~$\mu$m fluxes from W10.  The completeness at 4.5~$\mu$m is unclear due to the
nature of the 4.5~$\mu$m catalog (see W10 for details). 
\label{fig_color_mag} }
\end{figure}

Figure~\ref{fig_color_mag} shows a $K_S$ and 4.5~$\mu$m color--magnitude diagram for KIEROs 
and field galaxies.  The KIEROs are a relatively rare population among all sources detected at 4.5~$\mu$m,
and the selection of KIEROs lies very close to the detection limits at both $K_S$ and 4.5~$\mu$m.
At $K_S$, W10 provided a description of the completeness limits (e.g., the 
$K_S$ 30\% limit shown by the diagonal dashed line in Figure~\ref{fig_color_mag}).  On the other hand,
it is nontrivial to quantify the completeness at 4.5~$\mu$m, 
which depends on both the $K_S$ completeness and the very complex procedures used
by W10 to construct the 4.5~$\mu$m source lists.

%
%

%
%
\begin{figure}[t!]
\epsscale{1.0}
\plotone{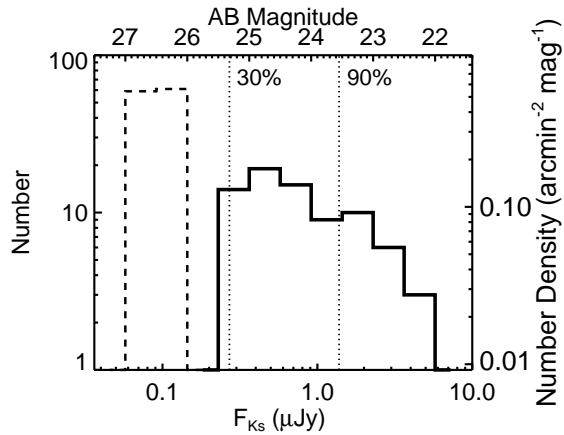}
\caption{$K_S$ number counts of KIEROs.
The solid histogram shows the number of sources detected in the
$K_S$ band at $>3\sigma$ in bins of 0.2 dex in flux.  The dashed histogram shows 
the same for sources with only $K_S$ upper limits ($1\sigma$).  The vertical dotted lines are the 90\% and 30\%
completeness limits of the WIRCam $K_S$ image derived by W10.
The actual completeness of the $K_S-4.5$~$\mu$m selection should be even lower (see text).
The number distributions shown here are \emph{not} corrected for completeness.
\label{fig_density}}
\end{figure}

Because of the above issues, we have decided to present the $K_S$ number counts of KIEROs 
(Figure~\ref{fig_density}), even though this is a 4.5~$\mu$m selected population.
In our survey there are no KIEROs brighter than 6 $\mu$Jy at $K_S$.  
The raw counts slightly flatten at $\sim 1$ $\mu$Jy
and drop rapidly below 0.3 $\mu$Jy, consistent with the $K_S$ detection
completeness.  Without considering the completeness at 4.5~$\mu$m,
simply correcting the counts with the $K_S$ completeness derived by W10 suggests that 
the slope of the counts does not change significantly down to a flux level of 
$\sim0.3$ $\mu$Jy.  A slope of $0.41\pm0.07$ in the logN-mag space
was fitted at $F_{Ks}>0.3$ $\mu$Jy using completeness corrected counts weighted by the Poisson errors.  
Comparing to the slope for the entire $K_S$ sample in the same flux range
($\sim0.2$, e.g., \citealp{maihara01,keenan10}), this is much steeper.  
This is likely a consequence of the higher redshifts of KIEROs (see next section).
Furthermore, the large number 
of $K_S$-undetected sources that meet our KIERO selection criterion (dashed histogram
in Figure~\ref{fig_density}) suggests that the counts may be still rising at a
$K_S$ flux of $\sim0.2$ $\mu$Jy.

\section{Redshift and Stellar Population}\label{sec_redshift}

\begin{figure}[t!]
\epsscale{1.0}
\plotone{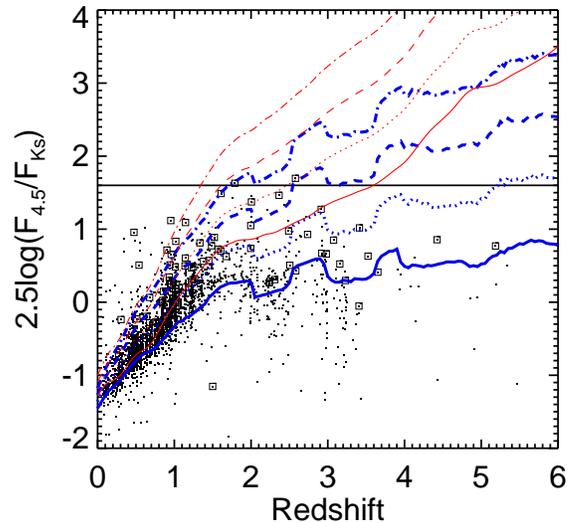}
\caption{$K_S$-4.5~$\mu$m color vs.\ spectroscopic redshifts.  Dots are 3$\sigma$ 4.5~$\mu$m 
sources in the catalog of W10 that are spectroscopically identified in Barger et al.\ (2008). Squares
show X-ray selected AGNs (soft or hard X-ray luminosities greater than $10^{42}$ erg s$^{-1}$).
Curves show colors derived from the models of elliptical galaxies (thin red curves, \citealp{cww80})
and starburst galaxies (thick blue curves, \citealp{kinney96}), reddened with the \citet{calzetti00}
extinction law with $A_V=0$ (solid), 1.0 (dotted), 2.0 (dashed), and 
3.0 (dash-dotted).
\label{fig_z_color} }
\end{figure}

To obtain a rough idea about the SEDs and redshifts of KIEROs,
we plot the color--redshift diagram of spectroscopically identified 4.5~$\mu$m sources in
Figure~\ref{fig_z_color}.  Most 4.5~$\mu$m sources have $K_S-4.5$~$\mu$m colors
bluer than 1.0.  Even AGNs, which tend to have unusual colors, 
are mostly bluer than 1.0.  In general, we only expect heavily extinguished ($A_V>2$)
galaxies at $z>2$ to be redder than 1.6.  At $z<2$, galaxies can be redder
than 1.6 in only the most extreme cases.  We do not expect to see many such galaxies.
To verify that KIEROs are mostly at $z>2$, we looked at the spectroscopic redshifts
of GOODS-N galaxies and carried out a photometric redshift analysis.

%
%

\subsection{Spectroscopic Redshift}

In the entire KIERO sample, only two sources are bright enough
in the optical and NIR to have spectroscopic redshifts in our
redshift survey \citep{barger08}.  Both are detected in the 2 Ms \emph{Chandra} 
image (source numbers 109 and 135 in the catalog of \citealp{alexander03}). 
One is the SMG GOODS 850-7 ($F_{1.4~\rm GHz}=58$ $\mu$Jy, $F_{850~\rm \mu m}=6.2$ mJy) 
at $z=2.578$.  The other is a $F_{1.4~\rm GHz}=83$ $\mu$Jy 
radio source at $z=1.790$.  This radio source is very close to the bright SMG 
GOODS 850-2 ($F_{850~\rm \mu m}=10.3$ mJy), with an angular separation of 
$\sim8\farcs6$ based on its SCUBA position (Wang et al.\ 2004; see also \citealp{barger00}).
However, our recent submillimeter interferometric observation (A.\ J.\ Barger et al., in preparation) 
shows that GOODS 850-2 is another faint $K_S$-band source that also enters our
KIERO sample.  

In addition to the two optical spectroscopic redshifts, an SMG KIERO, GOODS 850-5 
\citep{wang07}, has a plausible millimeter spectroscopic redshift of 4.042 based on a
line detection at 91.4 GHz that is likely a CO(4--3) transition \citep{daddi09b}.
The three spectroscopic redshifts span a redshift range that agrees with what would be expected 
based on Figure~\ref{fig_z_color}.  However, the sample size is too small and we need to
rely on photometric redshift fitting.

\subsection{Photometric Redshift}

We carried out a photometric redshift analysis on a subsample of 76 high S/N KIEROs whose ${K_S}$ fluxes are
greater than 0.2 $\mu$Jy.
We included the photometry in the $U$, \emph{HST} ACS, $J$, WFC3 F140W, $K_S$, and IRAC bands.
We adopted the $U$-band fluxes from \citet{capak04}.
For the ACS bands, we directly adopted the fluxes from the GOODS-N v2 catalog \citep{giavalisco04}.  
We measured $J$-band fluxes from a CFHT WIRCam image that contains 
$\sim27$ hr of total integration.  We obtained the $J$-band images from the public
archive. They were originally obtained by a Taiwanese team led by Lihwai Lin (2010, in preparation).  
We reduced and processed the $J$-band images in an identical manner to W10 for the
$K_S$-band images.  The \emph{HST} WFC3 F140W data will be described by 
A. Barger et al.\ (2011, in prep).

\begin{figure}[t!]
\epsscale{1.0}
\plotone{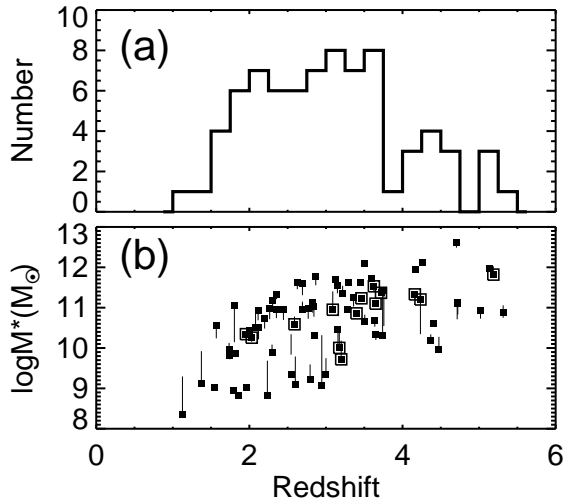}
\caption{Results of photometric redshifts for $F_{Ks}>0.2$ $\mu$Jy sources in the KIERO
sample.  A total of 76 KIEROs are included.  The top panel shows the distribution of the EAZY 
photometric redshifts.  The bottom panel shows the stellar masses fitted by
Hyperz at the EAZY redshifts using the models of \citet{bc03} (see Section~\ref{sec_stellar}).
The vertical bars show the maximum
and minimum stellar masses from various types of models, while the squares are the best-fit masses.
Large open squares denote AGNs identified in Section~\ref{sec_agn}, which may have problematic
photometric redshifts and masses.
\label{fig_photoz}} 
\end{figure}

For photometric redshift fitting, we tried two packages: the latest version of Hyperz 
\citep{bolzonella00}\footnotemark[6], and EAZY \citep{brammer08}.  
We found that EAZY generally produces better results and includes a necessary feature for
our studies (see below). Our primary photometric redshift results in this paper are thus based 
on EAZY.  We adopted the default set of SED templates
of \citet{br07}, provided in the EAZY package.  This set includes five templates ranging from
very blue and young galaxies with strong emission lines, to galaxies dominated by old stars.
We included an additional SMG-type dusty starburst model ($t=50$ Myr, $A_V=2.75$), 
also provided in the EAZY package. We refer to Brammer et al.\ (2008) for detailed descriptions on all 
these SED models. The templates here all already include certain amounts of extinction
in order to fit galaxy SEDs in deep surveys.  To account for very dusty sources in the KIERO sample, 
we further reddened these templates (including the already reddened SMG template) 
by $A_V=0$, 0.5, and 1.0 with the extinction law of \citet{calzetti00}.  
To account for the situation where there are more than two distinct stellar populations in a 
galaxy, we allowed for all linear combinations of the templates in the fitting.  In Section~\ref{sec_stellar}
we will show that the EAZY feature  of allowing for all linear combinations of templates is essential.  
Because a fitted SED is a combination of multiple templates, we are not able to quote an
extinction value for a source. 
In Figure~\ref{fig_photoz}(a) we show the distribution of 
the EAZY photometric redshifts.  

\footnotetext[6]{see also http://www.ast.obs-mip.fr/users/roser/hyperz/}

%
%

In order to see how reliable the fitting is, we ran EAZY on $\sim1500$ $F_{Ks}>0.2$ $\mu$Jy 
galaxies in the spectroscopic sample of Barger et al.\ (2008).  The results are shown in 
Figure~\ref{fig_photoz_compare}.  If we define $\Delta z$ as $(z_{\rm ph}-z_{\rm sp})/(1+z_{\rm sp})$,
then only 5.3\% of the galaxies have $|\Delta z| >0.2$.  After excluding these outliers, there 
is an rms dispersion of 0.05 and a median systematic offset of -0.01 in $\Delta z$.
We also made the same comparison on Hyperz results. The outlier fraction and dispersion in 
$\Delta z$ are both significantly worse than the EAZY ones.

%
%
\begin{figure}[h!]
\epsscale{1.0}
\plotone{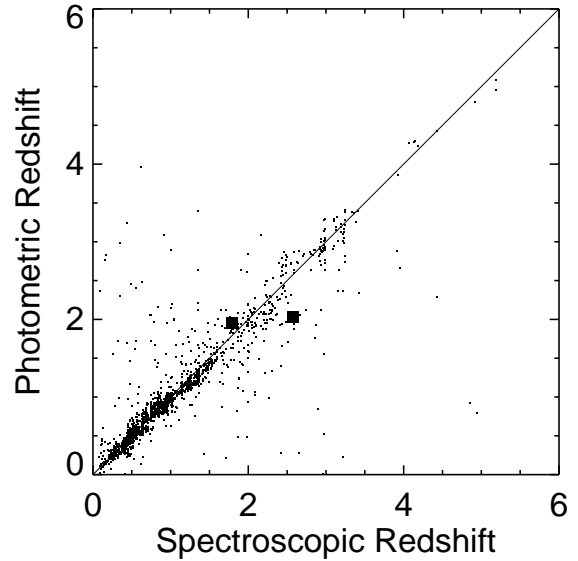}
\caption{Results of EAZY photometric redshifts for the $F_{Ks}>0.2$ $\mu$Jy sources (small dots)
in the spectroscopic sample of Barger et al.\ (2008).  A total of 1987 galaxies are included.  Large 
squares are the two KIEROs with spectroscopic redshifts.
\label{fig_photoz_compare}} 
\end{figure}

Although the results in Figure~\ref{fig_photoz_compare} look good, there is 
no guarantee that we can achieve the same on KIEROs.
The two KIEROs with spectroscopic redshifts at $z=2.578$ and 1.790 have 
EAZY photometric redshifts of 2.03 and 1.96, respectively (large squares in 
Figure~\ref{fig_photoz_compare}).  These are excellent, but
the sample is too small for us to comment on the overall quality.   We can also compare
the photometric redshifts on KIEROs that show continuum dropouts (Lyman breaks, see also 
Section~\ref{sec_lbg}). There are 15 $b$-dropouts in the KIERO sample, 6 of which are
bright enough at $K_S$ to have photometric redshifts. Their mean photometric redshift is $3.72\pm0.50$.  
There are 4 $v$-dropouts in the KIERO sample, 2 of which have photometric redshifts of 
5.02 and 4.71.
These are all consistent with the expected redshifts for the dropouts \citep[e.g.,][]{bouwens07}.
Finally, Figure~\ref{fig_sed} shows that the majority of the fitted SEDs reasonably represent the observed
SEDs, including various spectral breaks (the Lyman breaks, Balmer breaks, and 1.6 $\mu$m bumps),
except AGNs (Section~\ref{sec_agn}) with featureless power-law SEDs in the IRAC bands.  
It is worth noting that there are three
non-AGN data points at $z>4$ with unusually large stellar masses (Section \ref{sec_stellar}) of 
$\sim10^{12}M_{\sun}$. They are \#23, 24, and 38 in Figure~\ref{fig_sed}.
Their observed SEDs are also quite featureless in the IRAC bands, and all 
fade dramatically at $K_S$ and shorter wavebands, likely because of extinction. 
Although they are not identified as AGNs, we suspect that their photometric fittings are unreliable.
Except for AGNs and these few problematic cases, the EAZY photometric redshifts for the majority
of the sources seem reasonably good.  In Section~\ref{sec_submm} we present independent 
evidence suggesting that the redshift distribution in Figure~\ref{fig_photoz}(a) is approximately correct.

%
%
\begin{figure*}[t!]
\epsscale{1.15}
\plotone{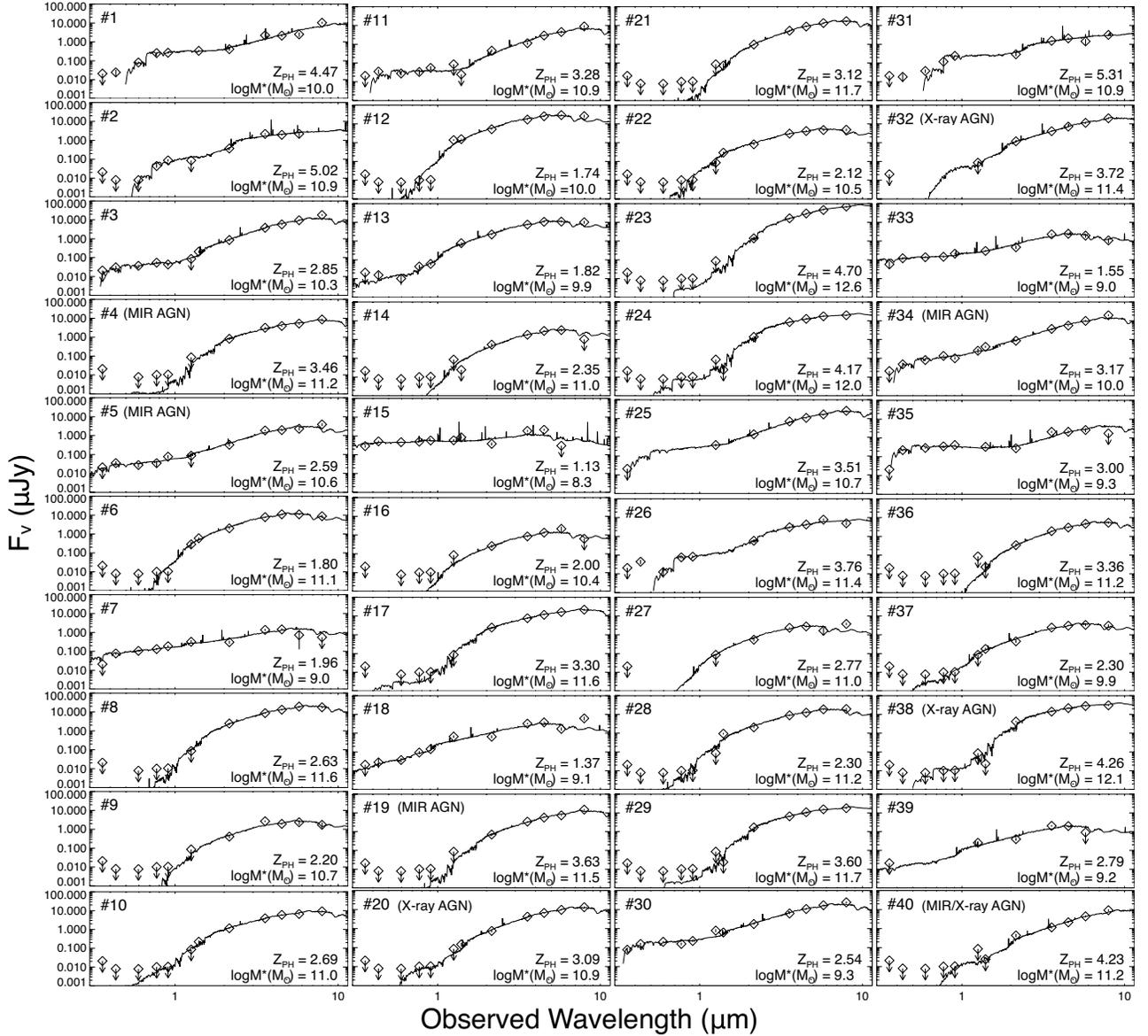}
\caption{Observed (diamonds) and fitted (curves) SEDs in the EAZY photometric redshift analysis.
Sources without symbols in the ACS and F140W bands are outside the image region.  Sources without 
symbols in the $J$ band do not have reliable $J$-band counterpart identifications.  AGNs identified
in Section~\ref{sec_agn} are indicated in the upper-left corners and have problematic photometric redshifts.
\label{fig_sed}} 
\setcounter{figure}{6}
\end{figure*}

%
%
\begin{figure*}[t!]
\epsscale{1.15}
\plotone{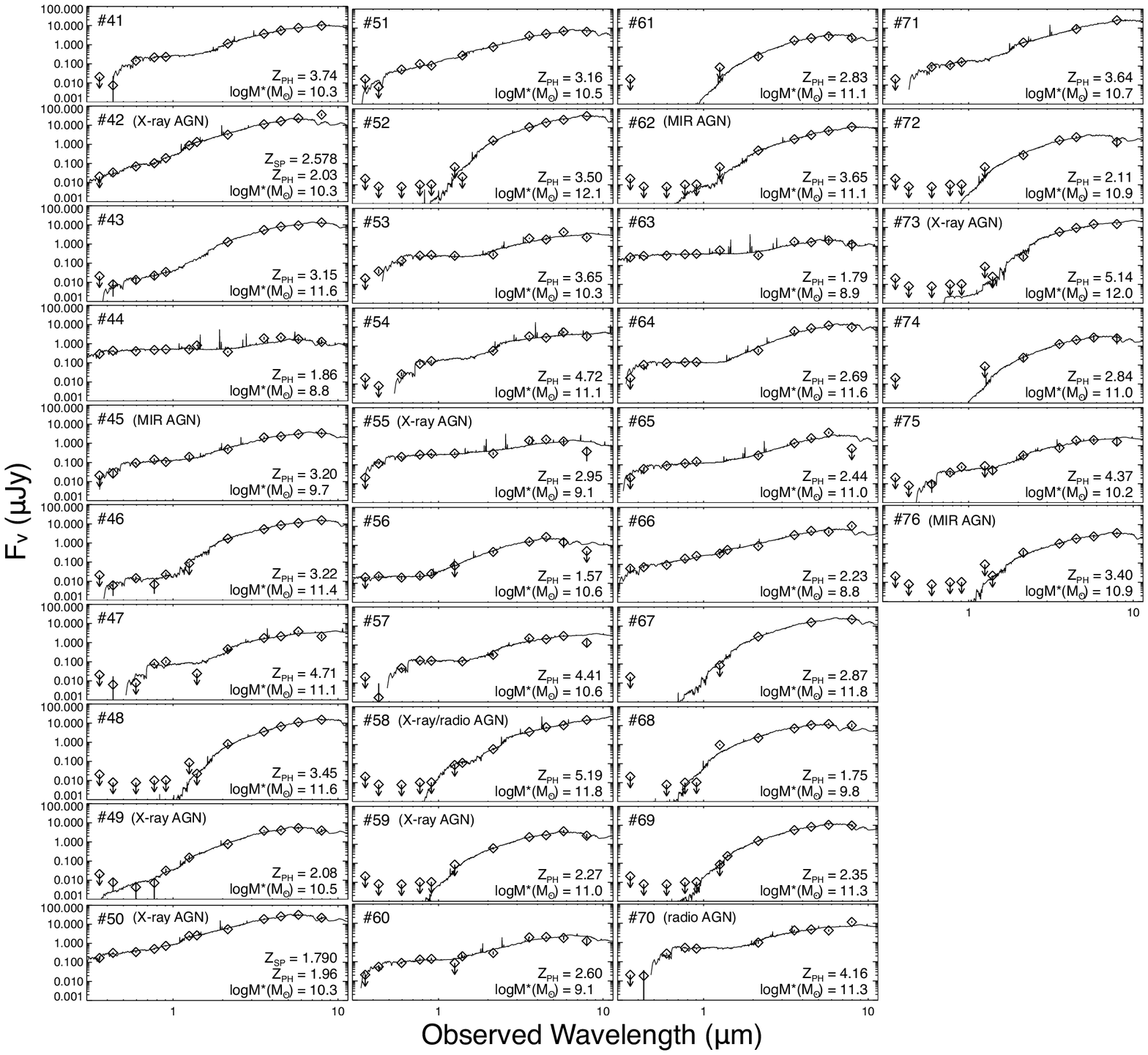}
\caption{Continued.} 
\end{figure*}

We did not attempt to study the redshifts of fainter sources ($F_{Ks}<0.2$ $\mu$Jy). 
Such sources often have robust photometry in only two IRAC bands, which is insufficient for good photometric 
redshifts.  One might expect that these fainter sources should have higher redshifts rather than lower
intrinsic luminosities.  However, we cannot test this with the current data.

\subsection{Stellar Population}\label{sec_stellar}

When testing the photometric redshift fitting, we found that many KIEROs have SEDs that cannot
be fitted with a single stellar population by Hyperz.  For example, on the above
mentioned six $b$-dropouts, Hyperz returns a mean redshift of $2.98\pm0.34$.  This is clearly too low,
a consequence of trying to fit the complex SEDs with single stellar populations.
This is a key reason for our adoption of EAZY, which allows for combinations of different SED templates.  
The most obvious cases of such KIEROs are sources 3, 11, 30, 35, 41, 53, and 70 in Figure~\ref{fig_sed}.
In such KIEROs, there are well-evolved old stellar populations, which produce the observed NIR luminosity, 
as well as unobscured ongoing star formation, which produces the strong rest-frame UV emission.  
This property is similar to that of objects selected with $z-3.6~\mu \rm m > 3.25$ (IEROs) by 
\citet{yan04} in the Hubble Ultra Deep Field (HUDF; \citealp{beckwith06}).  Interestingly, our stacking 
analyses in the radio and FIR (Section~\ref{sec_prop} and Table~\ref{tab_stack}) show that sources that 
are faint in the rest-frame UV have more intensive starbursts.  SFR estimates based on rest-frame UV
may miss the strong dust-hidden star formation on such optically faint sources.

Another interesting question to ask is whether or not KIEROs are massive galaxies.
Unfortunately, EAZY does not provide direct estimates of stellar masses.
To do this, we rely on Hyperz, which builds in the stellar population synthesis models of \citet{bc03}
and provides stellar mass estimates.  We forced Hyperz to fit at the EAZY redshifts.
We derived two sets of stellar masses by fitting to the full SEDs of the KIEROs, 
and by fitting to just the infrared SEDs in the $J$ and redder bands.
We found that neither of the two masses systematically favor high or low mass values.
For roughly 2/3 of the objects, these two masses agree within 0.4 dex. We thus believe 
that for most of the objects, the derived masses are robust. On the other hand, since many of
the KIEROs show two different stellar populations, it is better to just fit the rest-frame NIR parts
of the SEDs.  The mass-to-light ratios in the NIR are relatively insensitive
to both extinction and ongoing star formation. Nevertheless, we remind the reader that our mass
estimates are significantly limited by the fact that Hyperz can only fit the SEDs with
single stellar populations.

In Figure~\ref{fig_photoz}(b) we show the best fit stellar masses (squares) and the maximum and minimum 
masses (vertical bars) from the various templates of \citet{bc03}, based on just fitting the infrared SEDs.
For most KIEROs, the data only allow small ranges of stellar masses, a consequence
of similar mass-to-light ratios in the NIR.  They have large stellar masses of $10^{10} M_\sun$ to 
$\gtrsim10^{12} M_\sun$.  Some of the masses seem unusually large, especially at the high-redshift
end ($z>4$) where less photometric data points are available and there is a degeneracy between
photometric redshift, age, and extinction.  This is a fundamental limit of the current data.
We also note that there are 18 sources with stellar masses less than  $10^{10} M_\sun$,
giving a mass distribution that is somewhat disjoint from that of the other KIEROs.  These are all optically bright
sources with $F_{\rm F850LP}>0.2$ $\mu$Jy and have lower redshifts.  They are likely a less dusty subclass of
KIEROs.

In summary, the majority of KIEROs in the photometric redshift subsample are more massive
than $10^{10} M_\sun$.  In addition to the massive stellar populations, a large fraction of KIEROs 
show young stellar populations in the rest-frame UV. In the following sections of this
paper, we will focus on the star formation properties of KIEROs.

\section{Multiwavelength Properties and Stacking Analyses}\label{sec_prop}

\subsection{Radio Properties}\label{sec_radio}

In order to understand the star formation activities in KIEROs (albeit with AGN contamination), 
we studied their radio properties.
We used the Very Large Array (VLA) 1.4 GHz image and catalog in the Hubble Deep Field-North
(HDF-N) published by \citet{morrison10}. The image has a 5 $\sigma$ sensitivity of 20 $\mu$Jy
at the field center, and the catalog contains $\sim430$ sources in the GOODS-N region.
Among the 196 KIEROs, 22 are included in the catalog of \citet{morrison10}.  

\begin{figure}[h!]
\epsscale{1.0}
\plotone{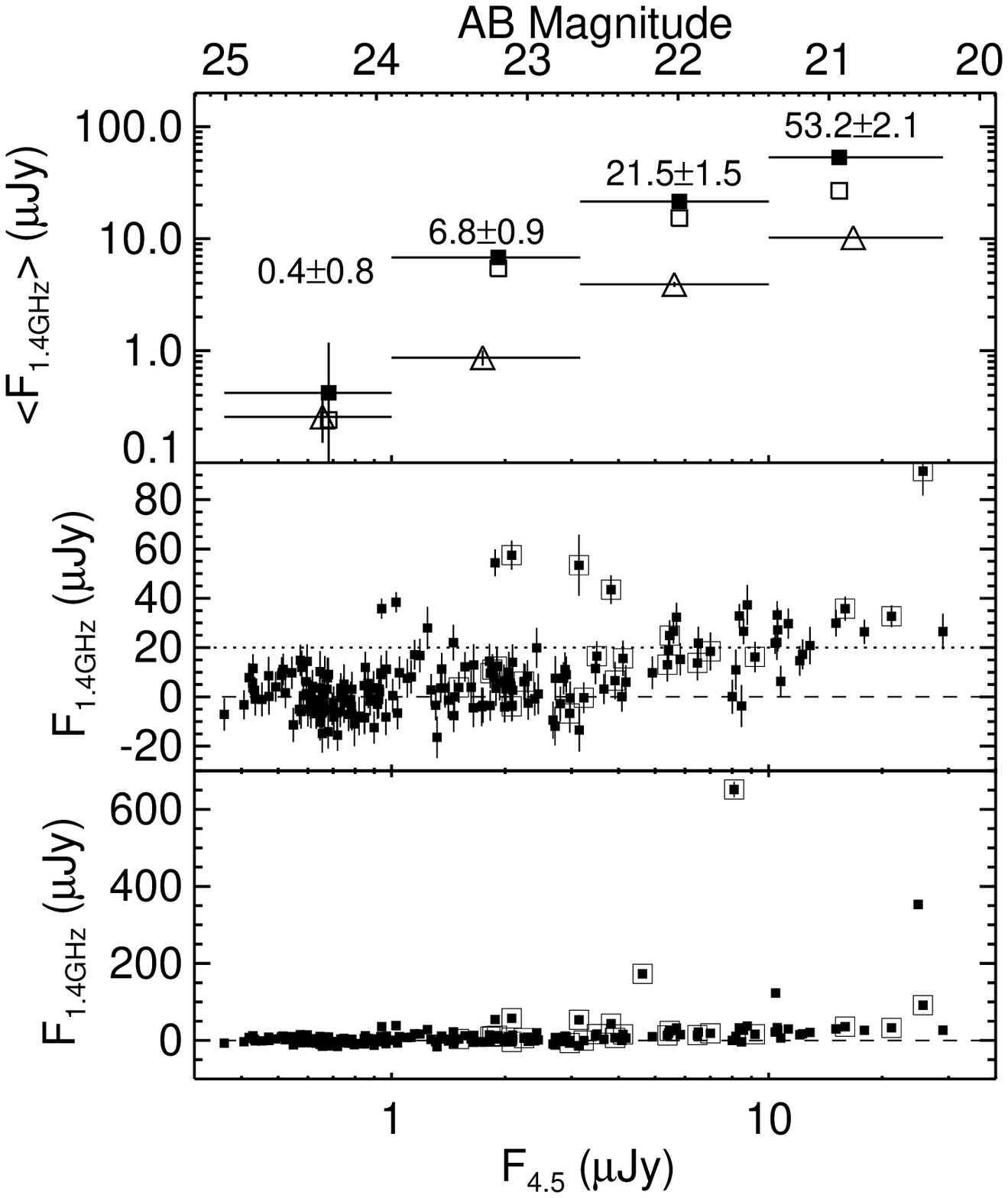}
\caption{Radio flux vs.\ 4.5~$\mu$m flux for KIEROs.  The bottom and middle panels show the
radio fluxes of individual sources. The middle panel is a blowup of the bottom panel.
Sources in the bottom and middle panels enclosed in large squares are AGNs identified 
in Section~\ref{sec_agn}.
The top panel shows the stacked radio fluxes in each 0.5 dex 4.5~$\mu$m flux bin.
Solid squares are the mean fluxes after excluding the $F_{1.4~\rm GHz}>600$ $\mu$Jy 
bright source.  The exact values are also shown near the symbols.
Open squares are the median fluxes, which further avoid influence from
$>200$ $\mu$Jy sources.  Triangles show the mean fluxes of the
entire 4.5~$\mu$m sample in the same flux ranges.  The dotted line corresponds to roughly $5\sigma$.  Stacking
of faint sources below this line is presented in the lower half of Table~\ref{tab_stack}.
\label{fig_radio_irac2}} 
\end{figure}

For sources not in the catalog of \citet{morrison10}, we measured their fluxes from the image 
with $3\arcsec$ apertures after correcting for the VLA primary beam and the band-width 
smearing.  We compared our fluxes on compact sources with those in the catalog of 
\citet{morrison10}, and they are fully consistent with each other.  The adopted aperture size is 
$2\times$ the VLA beam FWHM and corresponds to 20--25 kpc at $z=2$--5.  This size is slightly 
less optimized for detecting faint and compact sources, but it encloses most of the radio flux if the 
radio-emitting region in a high-redshift galaxy is offset from the position determined by NIR 
observations by a few kpc.  Such offsets are not uncommon in nearby interacting/merging systems.
The measured 1.4 GHz fluxes of the KIEROs are shown in
the bottom and middle panels of Figure~\ref{fig_radio_irac2}.

%
%

We can study the mean 1.4~GHz fluxes of KIEROs via a ``stacking analysis,'' in which
we measure radio fluxes with $3\arcsec$ apertures at the locations of KIEROs and average
the results.  First, we need to understand the bias and uncertainty in such stacked fluxes, 
for which we took a Monte Carlo approach. We randomly placed $3\arcsec$ apertures in the VLA 
image, calculated their mean flux, and repeated a large number of such measurements.
The mean flux measured in this way is an estimate of contamination from physically unrelated nearby
sources.  We refer to this as confusion.  We excluded flux values greater than 600 $\mu$Jy 
to avoid being biased by the very bright source, which is likely an AGN 
(also see Section~\ref{sec_x}). In the IRAC region, the mean 1.4~GHz flux is 0.41 $\mu$Jy
for the random positions in the Monte Carlo simulations.
We subtracted this value from the mean flux of each KIERO sample.  

We treated the dispersion between various measurements of the mean flux of random apertures as 
noise in the mean, which has contributions from the receiver noise and from the confusion noise.  
We varied the number of random apertures that we placed in the image each time, and we found that even 
with source densities up to $10^6$ deg$^{-2}$, the measured noise in
the mean precisely scales with the square-root of the source number.
In the IRAC region, such noise is 8.7 $\mu$Jy per source. 
This value is larger than the nominal
4 $\mu$Jy noise of the image, a consequence of confusion caused by bright sources.  On the other hand, 
the average fluxes we determine are considerably below the conventional
single-source detection limit.

We present our major stacking analysis results in Table~\ref{tab_stack}.
For the entire sample of 196 KIEROs, excluding the $>600$ $\mu$Jy one, the 
mean 1.4~GHz flux is $9.9 \pm 0.6$ $\mu$Jy, and the median is 3.07 $\mu$Jy.
For comparison, in the 4.5~$\mu$m sample there are 8977 sources with 
$F_{4.5~\rm \mu m}>1$ $\mu$Jy. 
Their mean 1.4 GHz flux is $4.7\pm0.1$ $\mu$Jy, and their median flux is 2.46 $\mu$Jy.  
After considering the fact that most of the 4.5~$\mu$m
sources are at $z<2$ while most of the KIEROs are at $z>2$, it is clear that our 
$K_S-4.5$~$\mu$m $>1.6$
selection picks up a much more radio luminous population.

%
%
\begin{deluxetable*}{lrcccrcccrcc}[h!]
\tablewidth{0pt}
\tablecaption{Stacking Results in the Radio, Millimeter, and Submillimeter \label{tab_stack}}
\tablehead{\colhead{Samples} & \multicolumn{3}{c}{1.4 GHz Stacking (0.06 deg$^2$)} && \multicolumn{3}{c}{1100 $\mu$m Stacking (0.06 deg$^2$)} && \multicolumn{3}{c}{850 $\mu$m Stacking (0.028 deg$^2$)} \\
\cline{2-4}  \cline{6-8} \cline{10-12}
& \colhead{$N$\tablenotemark{a}} 	& \colhead{$\langle F \rangle$\tablenotemark{c}} 	& \colhead{EBL\tablenotemark{b}} & &
\colhead{$N$\tablenotemark{a}} 		& \colhead{$\langle F \rangle$} 					& \colhead{EBL\tablenotemark{b}} & &
\colhead{$N$\tablenotemark{a}} 		& \colhead{$\langle F \rangle$} 					& \colhead{EBL\tablenotemark{b}} \\
& & \colhead{($\mu$Jy)} & \colhead{(mJy deg$^{-2}$)} &&
& \colhead{(mJy)} & \colhead{(Jy deg$^{-2}$)} &&
& \colhead{(mJy)} & \colhead{(Jy deg$^{-2}$)}}
\startdata
$F_{4.5~\rm \mu m}>1$ $\mu$Jy 	& 8977 &	$4.7\pm0.1$ & 	$706\pm15$ 	&& 8984 & $0.090\pm0.01$ &	$13.4\pm1.8$ 		&& 4551 &$0.14\pm0.04$ & $22.7\pm6.4$ \\
KIERO 						& 195    &	$9.9\pm0.6$ & 	$32.2\pm2.0$	&& 196 & 	$0.50\pm0.08$ &    	$1.64\pm0.26$		&& 87 &	$1.44\pm0.28$ & $4.47\pm0.88$ \\
KIERO, $z<3.0$ 				& 38    &	$22.7\pm1.4$ &$14.4\pm0.9$	&& 38 & 	$0.56\pm0.18$ & 	$0.36\pm0.12$		&& 18 & 	$2.48\pm0.62$ & $1.59\pm0.40$ \\
KIERO, $z>3.0$ 				& 37 & 	$21.0\pm1.4$ &$12.9\pm0.9$	&& 38 & 	$1.57\pm0.18$ & 	$1.00\pm0.12$ 	&& 14 &	$3.19\pm0.71$ & $1.59\pm0.35$ \\
KIERO, ACS           				& 112 &	$6.0\pm1.0$ & 	$11.2\pm1.5$	&& 112 & 	$0.23\pm0.11$ &  	$0.42\pm0.20$		&& 52 &	$0.71\pm0.37$	& $1.32\pm0.68$ \\
KIERO, non-ACS    				& 61	&	$16.2\pm1.1$ &$16.7\pm1.1$	&& 62 & 	$0.94\pm0.14$	& 	$0.97\pm0.14$		&& 32 &	$2.59\pm0.47$ & $2.96\pm0.54$ \\
\hline
& \multicolumn{3}{c}{$F_{1.4~\rm GHz}<20~\mu$Jy Sources} & & \multicolumn{3}{c}{$F_{1100~\rm \mu m}<3$ mJy Sources} & & \multicolumn{3}{c}{$F_{850~\rm \mu m}<6$ mJy Sources} \\
\hline
$F_{4.5~\rm \mu m}>1$ $\mu$Jy 	& 8282 &	$1.76\pm0.07$ & 	$242\pm10$ 	&& 8849 & $0.062\pm0.011$ &	$9.2\pm1.6$ 		&& 4446 &$0.092\pm0.037$ & $14.5\pm5.8$ \\
KIERO 						& 165    &	$2.6\pm0.5$ & 		$7.14\pm1.45$	&& 177 & 	$0.14\pm0.08$ &    	$0.41\pm0.23$		&& 78 &	$0.65\pm0.28$ & 	$1.82\pm0.77$ \\
KIERO, $z<3.0$ 				& 27    &	$2.4\pm1.3$ &		$1.07\pm0.59$	&& 34 & 	$0.20\pm0.18$ & 	$0.12\pm0.10$		&& 15 & 	$0.90\pm0.63$ &	$0.48\pm0.34$ \\
KIERO, $z>3.0$ 				& 23 & 	$8.6\pm1.4$ &		$3.31\pm0.54$	&& 28 & 	$0.72\pm0.20$ & 	$0.33\pm0.09$ 	&& 9 &	$0.78\pm0.81$ & 	$0.25\pm0.26$ \\
KIERO, ACS           				& 102 &	$1.1\pm0.7$ & 		$1.93\pm1.14$	&& 107 & 	$0.09\pm0.10$ &  	$0.16\pm0.18$		&& 49 &	$0.31\pm0.35$	&	$0.54\pm0.61$ \\
KIERO, non-ACS    				& 48	&	$5.0\pm1.0$ &		$4.00\pm0.80$	&& 51 & 	$0.21\pm0.14$	& 	$0.18\pm0.11$		&& 26 &	$1.16\pm0.48$ &	$1.08\pm0.45$ 

\enddata
\tablenotetext{a}{$N$ is the number of sources included in each sample at the corresponding wavelength.}
\tablenotetext{b}{EBL is the surface brightness contribution to the extragalactic background light.}
\tablenotetext{c}{In the radio stacking, we always exclude sources brighter than 600 $\mu$Jy to avoid bias from radio AGNs (see text).  Such bright sources bias
the stacked fluxes of both the 4.5~$\mu$m population the KIERO population by $\sim30\%$.}
\end{deluxetable*}

The high S/N of the mean 1.4~GHz flux for KIEROs allows us to further break down the
sample.  In the upper panel of Figure~\ref{fig_radio_irac2}, we show the
mean 1.4 GHz flux of the KIEROs as a function of 4.5~$\mu$m flux (filled squares).  
(For comparison, we show the same for the 8977 $F_{4.5~\rm \mu m}>1$ $\mu$Jy sources 
with open triangles.)
There is a strong linear correlation between the mean radio flux and the 4.5 $\mu$m flux.
In our KIERO sample the mean $3.6~\mu \rm m - 4.5~\mu m$ color is 0.19, 
which corresponds to a spectral index of $\alpha=0.78$ ($f_\nu\propto\nu^{-\alpha}$).  
This is similar to the radio spectral index of normal galaxies (0.7--0.8), implying similar 
$K$-corrections in the radio and 4.5~$\mu$m. Therefore, the correlation 
could suggest an approximately linear relationship between average SFR measured 
from the radio power and stellar mass measured from the rest-frame NIR.
However, in Section~\ref{sec_sfr}, we show that the mean radio fluxes in two
redshift bins between $z=2$ and 4 do not obviously depend on the stellar masses.
So the linear relationship between radio and 4.5 $\mu$m fluxes observed here is 
more likely to be just an apparent effect of redshift.

%
%
\begin{figure}[h!]
\epsscale{1.1}
\plotone{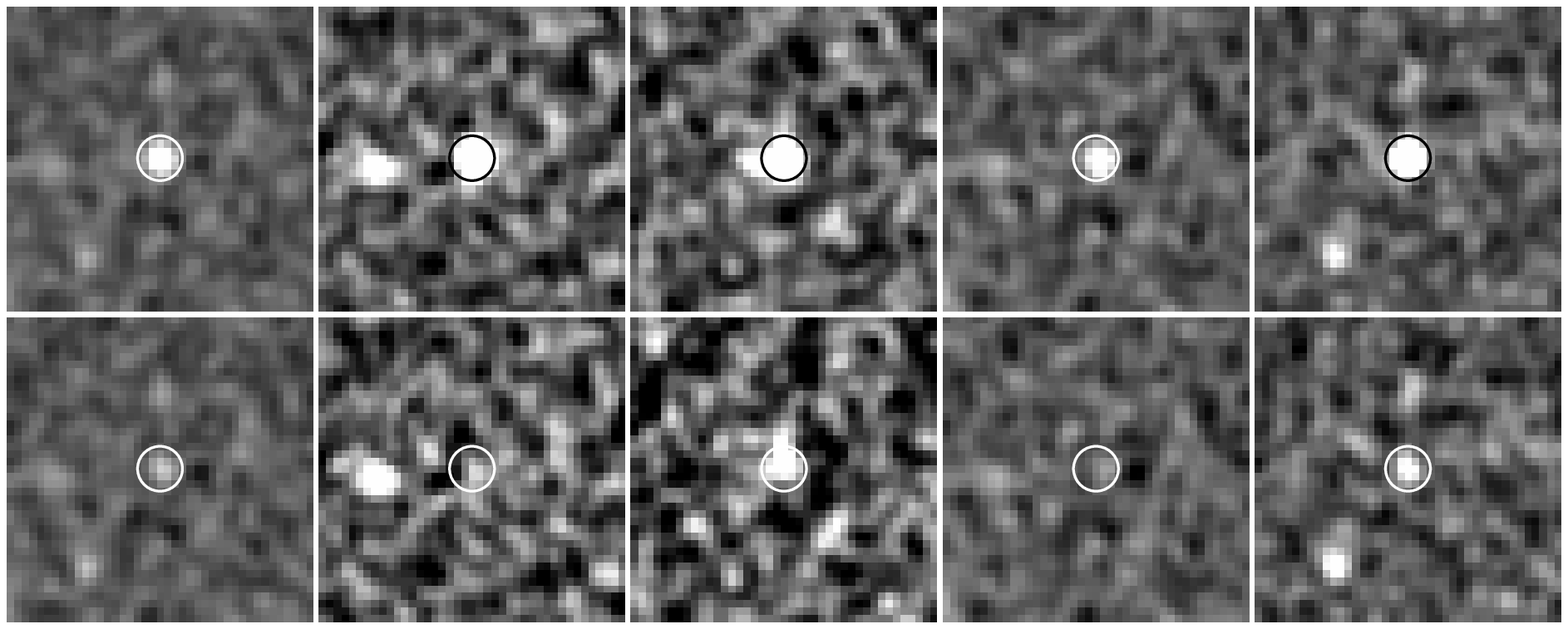}
\caption{Mean radio images of the KIERO subsamples in Table~\ref{tab_stack}.
The top row shows the results of $F_{1.4~\rm GHz}<600~\mu$Jy sources, and
the bottom row shows the results of $F_{1.4~\rm GHz}<20~\mu$Jy sources.
From left to right are all sources fainter than the above flux limits, $z<3$ sources, $z>3$ sources,
ACS detected sources, and ACS undetected sources.  Each panel has a size of $20\arcsec$.
The small circles indicate our $3\arcsec$ flux apertures.  The grayscale ranges in all panels 
are $-1.5$ to $3$ $\mu$Jy beam$^{-1}$.
\label{fig_stack_radio_image}} 
\end{figure}

In Figure~\ref{fig_stack_radio_image} we show the stacked radio images of various KIERO 
subsamples.  We start by discussing the top row, where we have applied a limit of 
$F_{1.4~\rm GHz}<600~\mu$Jy.  The first thumbnail shows the stack of all the 
$F_{1.4~\rm GHz}<600~\mu$Jy sources.
We then stacked the radio fluxes of the photometric redshift subsample of 76 KIEROs.  
These sources have, on average, higher 1.4~GHz fluxes than other KIEROs, since they are 
selected to be brighter in the NIR, and the radio flux correlates with the NIR flux. We find that 
the KIEROs at $z<3$ (second thumbnail) and the KIEROs at $z>3$ (third thumbnail) 
have similar mean radio fluxes.  The observed radio flux for a 
given source has a redshift dependence of $F_{\rm radio} \propto (1+z)^{1-\alpha}/d_L(z)^2$, 
where $d_L$ is the luminosity distance.  Using this relation and the measured mean radio 
fluxes, we find that KIEROs at $z>3$ are $\sim2\times$ more radio luminous than
KIEROs at $z<3$ (see also, Section~\ref{sec_sfr} and Table~\ref{tab_sfr}).  This may be 
because the 4.5 $\mu$m selection is only sensitive to more luminous systems at higher redshifts.

We also stacked the radio fluxes of the ACS-detected (fourth thumbnail)
and the ACS-undetected (fifth thumbnail) KIEROs.
There is a striking \emph{anti-correlation} between the radio fluxes and the optical fluxes of KIEROs.
The ACS-undetected subsample is, on average, $2.5\times$ brighter than the ACS-detected subsample 
in the radio.  The same anti-correlation is also clearly observed when using 850 and 1100~$\mu$m 
fluxes rather than radio fluxes (see next section).
This shows that the most active star forming galaxies in the universe are deeply hidden in dust
and can hardly be traced with rest-frame UV observations, even with the extremely deep ACS observations
in the GOODS-N.

It is also important to ask whether the above results are general properties of the entire 
KIERO population, or are biased by a few bright sources.  In order to investigate this,
we removed sources brighter than 20 $\mu$Jy in the radio (dotted line in Figure~\ref{fig_radio_irac2}) 
and repeated the same stacking procedures.  
We adopted the same 20 $\mu$Jy cut in the Monte Carlo determinations of the 
confusion effect. The stacking results 
are listed in the lower half of Table~\ref{tab_stack}.  As expected, all the mean fluxes decrease,
but they are still greater than at least 1.5$\sigma$.  The general pattern is similar: the KIERO population
is, on average, brighter in the radio than the 4.5 $\mu$m sources; and the ACS-undetected KIEROs 
are brighter in the radio than the ACS-detected KIEROs.  We show some of the results in the bottom row
of Figure~\ref{fig_stack_radio_image}, where we have applied the $F_{1.4~\rm GHz}<20~\mu$Jy limit.
The thumbnail images correspond to all of the KIEROs below this limit (first), only the $z<3$ sources (second),
only the $z>3$ sources (third), only the ACS-detected sources (fourth), and only the ACS-undetected sources
(fifth).

In Figure~\ref{fig_radio_irac2} we see that nearly all (85 out of 86) sources with $F_{4.5~\rm \mu m}<1~\mu$Jy
are below the 20 $\mu$Jy radio flux cut.  Their stacked radio flux is $0.42\pm0.76$ $\mu$Jy, 
or $0.18\pm0.73$ $\mu$Jy after removing the one $>20~\mu$Jy source.  Both are consistent
with a null detection. On the other hand, if we look at sources with $F_{4.5~\rm \mu m}>1~\mu$Jy,
the positive signal in their radio fluxes is very clear.  In the $F_{4.5~\rm \mu m}=1$--3.2~$\mu$Jy bin,
the mean radio flux is $6.76\pm0.90$ $\mu$Jy for 65 sources, or $3.44\pm0.88$ after excluding the 
six $>20~\mu$Jy sources.

We conclude that the majority of the NIR bright ($F_{4.5~\rm \mu m}>1~\mu$Jy) KIEROs are luminous in the radio,
especially the optically faint sub class.  This is true even if we remove radio detected KIEROs from the sample.  
On the other hand, the NIR faint ($F_{4.5~\rm \mu m}<1~\mu$Jy) KIEROs 
(43\% of the total KIERO population) do not have detectable radio emission.  In Section~\ref{sec_sfr},
we will use the stacked radio fluxes to study the star formation properties of KIEROs.

\subsection{Millimeter and Submillimeter Properties}\label{sec_submm}

There exist numerous millimeter and submillimeter continuum surveys and follow-up studies
in the GOODS-N (e.g., \citealp{hughes98,barger00,borys03}; Wang et al.\ 2004; 
\citealp{wang06,pope06,perera08,greve08,chapin09}).  
Here we considered two source catalogs, our SCUBA 850~$\mu$m 
catalog (Wang et al.\ 2004) and the AzTEC 1100~$\mu$m catalog \citep{perera08}, both with
nearly complete multiwavelength identifications \citep{pope06,chapin09}.  To be conservative, 
we only considered robust sources ($4\sigma$)
with unambiguous 1.4~GHz identifications that we agree with, or with submillimeter 
interferometric identifications.  

There are 28 850~$\mu$m and 1100~$\mu$m selected sources with such identifications in 
the IRAC area, 14 of which pass our $K_S-4.5$~$\mu$m $>1.6$ criterion.  
In addition, we found that several unidentified millimeter and submillimeter sources have nearby KIEROs
within a few arcsec, sometimes multiple ones.  The high fraction of submillimeter sources is consistent with our 1.4~GHz
finding that we are selecting active star forming galaxies at high redshift with our KIERO selection.

We also stacked the millimeter and submillimeter fluxes of KIEROs.  
Here we used the 850~$\mu$m image of Wang et al.\ (2004) and the 1100~$\mu$m image
of \citet{perera08}.  The stacking is similar to that in the radio.  
However, the caveat here is a possible upward bias caused by the coarse beams in the millimeter and submillimeter.
If the stacked population has strong clustering at scales smaller or comparable
to the beam, then the stacked flux would be positively biased.  This was discussed by
Wang et al.\ (2006) and \citet{serjeant08} and recently demonstrated by \citet{marsden09}.
Marsden et al.\ performed a stacking analysis using the deep Very Large Telescope (VLT) NIR sample 
of \citet[$5\times10^5$ sources per deg$^2$]{grazian06}
in the GOODS-S and the BLAST 250--500~$\mu$m images 
(beam size $0\farcm6$--$1\arcmin$).  Their stacked far-IR (FIR) fluxes from the VLT NIR galaxies exceeded
the FIR extragalactic background light (EBL) by factors of 2.5--3\footnotemark[6].  The authors attributed this to small-scale 
clustering of the NIR selected galaxies, which shows a $\sim100\%$ excess in the number of
galaxies at scales of $\sim1\arcmin$.

\footnotetext[6]{Our group (Wang et al.\ 2006) did not find a stacked SCUBA 850 $\mu$m flux
that exceeds the EBL limit, although we used extremely deep and dense optical and NIR samples.
The difference between our results and that in \citet{marsden09} could be attributed to 
the dramatic difference in beam sizes of SCUBA and BLAST, or issues in the 
methodology adopted by either group.}

We performed a similar clustering analysis and found that our KIERO population only shows a 7\% and 9\% 
excess of sources within $15\arcsec$ (SCUBA beam FWHM) and $18\arcsec$ (AzTEC
beam FWHM), respectively.  This is much lower than the $\sim100\%$ excess of the VLT NIR
sample at the size scales of the BLAST beams.  Thus, we conclude that stacking analyses 
of the GOODS-N SCUBA and AzTEC images using the KIERO population should not be 
biased by more than a few percent by small-scale clustering.  (For comparison, we made
similar tests on the entire $K_S$-band sample.  There is a $\sim30\%$ and $\sim40\%$ excess
of sources at size scales of $15\arcsec$ and $18\arcsec$, respectively.)  
After integrating the excess amounts convolved with the beam profiles, we found that these
would bias the stacking results by up to 10\%.  This is likely an upper limit, since 
not all clustered objects are millimeter or submillimeter sources, as pointed out by \citet{serjeant08}.
We therefore conclude that bias caused by small-scale clustering in our sample is unlikely
to affect our millimeter and submillimeter stacking results.

Instead of aperture photometry, the 850 and 1100~$\mu$m fluxes were measured with
weighted point-source filters.  The filtering and weighting schemes are described in Wang et al.\ (2004) and \citet{perera08}.
We measured the noise (sky, instrumental, and confusion noises) in the weighted fluxes 
and the flux bias (caused by \emph{random} nearby sources) in a way identical 
to that for our radio stacking.  Our noise measurements are 2.99 and 1.13 mJy
per source for SCUBA and AzTEC, respectively.  The flux biases are both zero, because of the zero 
sums of the beams. In our SCUBA image, there are two negative 50\% sidelobes 
produced by the secondary chopping, producing a beam that has a zero total power.
The PSF of AzTEC is more complex but the filtering function adopted by \citet{perera08}
also has a zero sum.

%
%
\begin{figure}[h!]
\epsscale{1.0}
\plotone{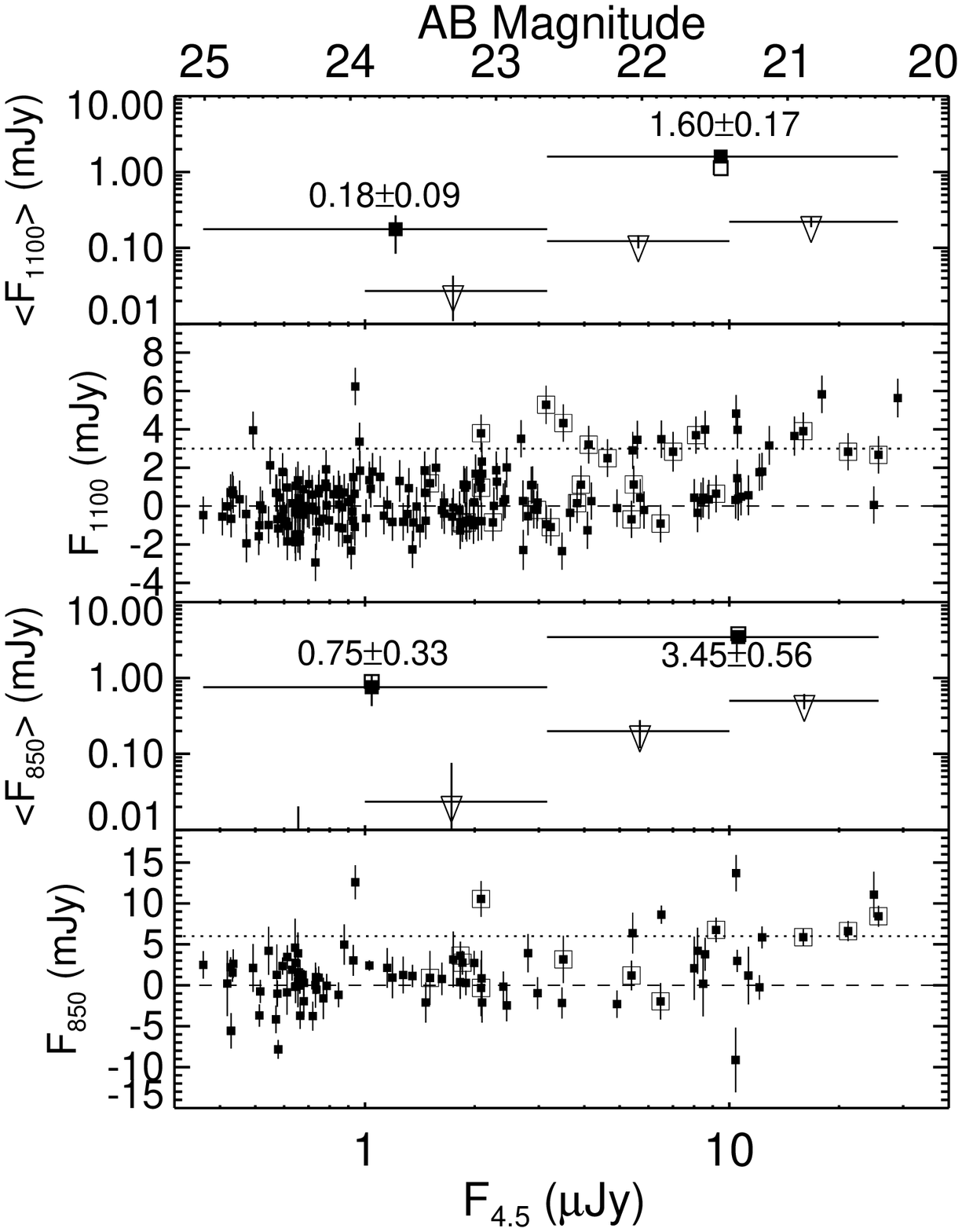}
\caption{Millimeter and submillimeter fluxes vs.\ 4.5~$\mu$m flux for KIEROs.
From the top we show stacked 1100~$\mu$m fluxes, 1100~$\mu$m fluxes of individual sources,
stacked 850~$\mu$m fluxes, and 850~$\mu$m fluxes of individual sources.
Individual sources enclosed in large squares are AGNs identified in Section~\ref{sec_agn}.
In the panels for stacked fluxes, solid squares show error-weighted stacked 
fluxes within bins of 1 dex in $F_{4.5~\rm \mu m}$ for the KIEROs.
The exact values are shown near the symbols.  Open squares show median 
fluxes for the same KIEROs in the same bins.  For comparison, we show with 
downward triangles stacked fluxes from the 4.5~$\mu$m selected sample
in 0.5 dex bins. Unlike the case in Figure~\ref{fig_radio_irac2}, these are upper limits 
and may be overestimated by a few to $\sim10$ percent because
of small-scale clustering (see text).  The 4.5~$\mu$m population is not detected at
both 850 and 1100~$\mu$m at $F_{4.5~\rm \mu m}<1$ $\mu$Jy, despite the great numbers of
sources there.  The dotted lines correspond to roughly $5\sigma$.  Stacking results
for faint sources below these lines are presented in the lower half of Table~\ref{tab_stack}.
\label{fig_mm_irac2}} 
\end{figure}

In Figure~\ref{fig_mm_irac2} we present the measured millimeter and submillimeter 
fluxes of the individual sources in the second and fourth panels, respectively, and the
stacked fluxes in the first and third panels, respectively.  We summarize the
stacking results in Table~\ref{tab_stack}.
As was the case in the radio, the millimeter and submillimeter fluxes seem to
correlate with the 4.5~$\mu$m flux.  Unfortunately, the S/N here does not
allow us to meaningfully break down the sample to further study this property.
When compared to the entire 4.5~$\mu$m sample (open downward triangles in 
Figure~\ref{fig_mm_irac2}), the KIERO selection picks up brighter objects, on average, 
which is also similar to the case in the radio.

We stacked the millimeter and submillimeter fluxes in the photometric redshift subsample.
At both 850~$\mu$m and 1100~$\mu$m, the mean fluxes from $z>3$ sources are higher than those 
from $z<3$ sources.  It is well know that the observed flux in the millimeter and submillimeter
is not a strong function of redshift at $z\sim1$ to 10 for a fixed luminosity
\citep{blain93}.  The weighted-combined 850 and 1100~$\mu$m result indicates that KIEROs at 
$z>3$ are $1.5\pm0.4$ times more luminous than KIEROs at $z<3$ are, on average.
This is consistent with the result from the radio stacking.  

As in the radio stacking analyses, here we also investigate the contribution from faint sources.
We removed 1100 $\mu$m data above 3 mJy and 850 $\mu$m data above 6 mJy (dotted lines in
Figure~\ref{fig_mm_irac2}), and repeated all the stacking procedures.  The results are listed in
the lower half of Table~\ref{tab_stack}.  The general trend here is very similar to that in the radio,
but with worse S/N.

We compared the millimeter and submillimeter and 1.4 GHz stacked fluxes and
obtained rough redshift estimates.  This ``millimetric redshift'' method was first
published by \citet{carilli99} and independently developed by \citet{barger00}.  
Here we adopted the formula derived by \citet{barger00},
\begin{equation}
z = 0.98(S_{850}/S_{1.4~\rm GHz})^{0.26}-1.
\end{equation}
In deriving this equation, Barger et al.\ assumed a radio spectral index of
$\alpha=0.8$ and a submillimeter dust emissivity of $\beta=1.0$,
and they used Arp 220 for flux normalization. 
Using our mean 850~$\mu$m flux, our mean 1100~$\mu$m flux extrapolated 
to 850~$\mu$m assuming
$\beta=1.0$, and our mean 1.4 GHz flux for $z<3.5$, we find an 
error-weighted millimetric redshift of 
$z\sim2.4$.  Using the stacked fluxes on undetected sources, we find a consistent 
redshift of 2.6.  Despite the large uncertainty of this method, the results are
remarkably close to what would be expected from the photometric redshift
distribution in Figure~\ref{fig_photoz}(a), which is unlikely to be a coincidence.
The agreement between these two entirely independent redshift estimates
strongly suggests that our photometric redshift estimates and
stacked fluxes are fairly reliable.

\subsection{X-Ray Properties}\label{sec_x}

Seven of the 196 KIEROs are in the 2~Ms \emph{Chandra} catalog
\citep{alexander03}.  These include the two spectroscopic KIEROs at $z=1.790$ 
and 2.578.  If we assume $z>2$ for the remaining five and an X-ray photon index of 
$\Gamma=1.8$ \citep[e.g,][]{barger07}, then all seven have either rest-frame hard or
soft X-ray luminosities exceeding $10^{42}$ erg s$^{-1}$ and hence are
X-ray AGNs.  One of the seven sources is the very bright 627 $\mu$Jy radio
source (source 58 in Figure~\ref{fig_sed}).  This extremely large radio flux and its 
$F_{1100~\mu\rm m}$ upper limit (1 mJy) 
cannot be explained by any known starburst SED at $z>0.5$.  Thus, its radio emission
is almost certainly powered entirely by an AGN.
This justifies the exclusion of this source in our radio stacking analysis,
since our main goal is to measure SFRs.   

\begin{figure}[h!]
\epsscale{1.0}
\plotone{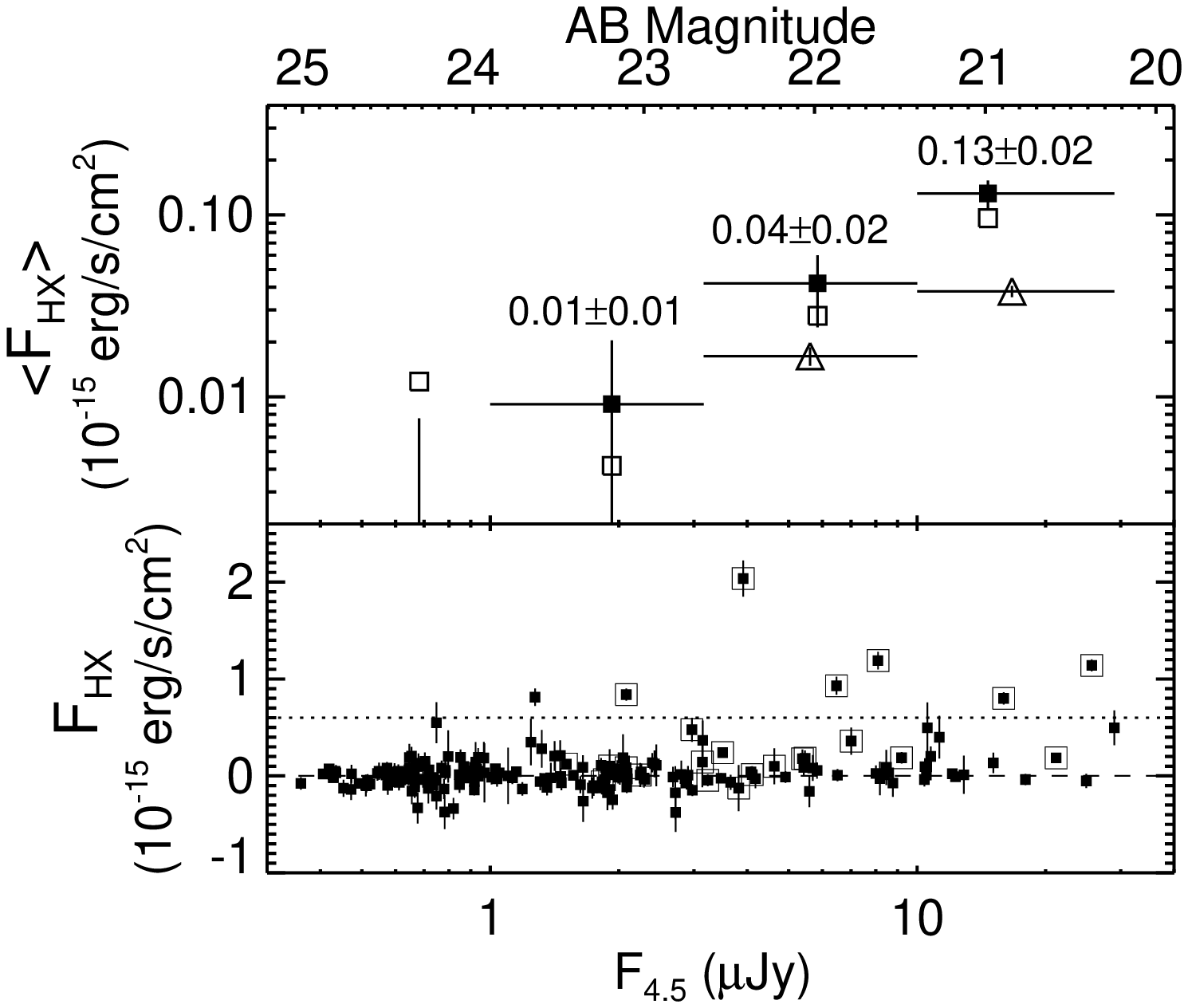}
\caption{Hard X-ray (2--8 keV) flux vs.\ 4.5~$\mu$m flux for KIEROs.  The bottom panel shows 
the X-ray fluxes of individual sources.  Sources enclosed in large squares are AGNs identified 
in Section~\ref{sec_agn}.  The top panel shows the stacked X-ray fluxes in each 
0.5 dex 4.5~$\mu$m flux bin, \emph{only for sources below the dotted line}, which 
roughly corresponds to $5\sigma$.  Solid squares are the mean fluxes for KIEROs.  
The exact values are also shown near the symbols.
Triangles show the mean fluxes of the
entire 4.5~$\mu$m sample in the same flux ranges.  
In the two lower 4.5~$\mu$m flux bins, both the KIERO sample and the 4.5~$\mu$m selected 
sample are not detected.
\label{fig_hx}} 
\end{figure}

\begin{figure}[h!]
\epsscale{1.0}
\plotone{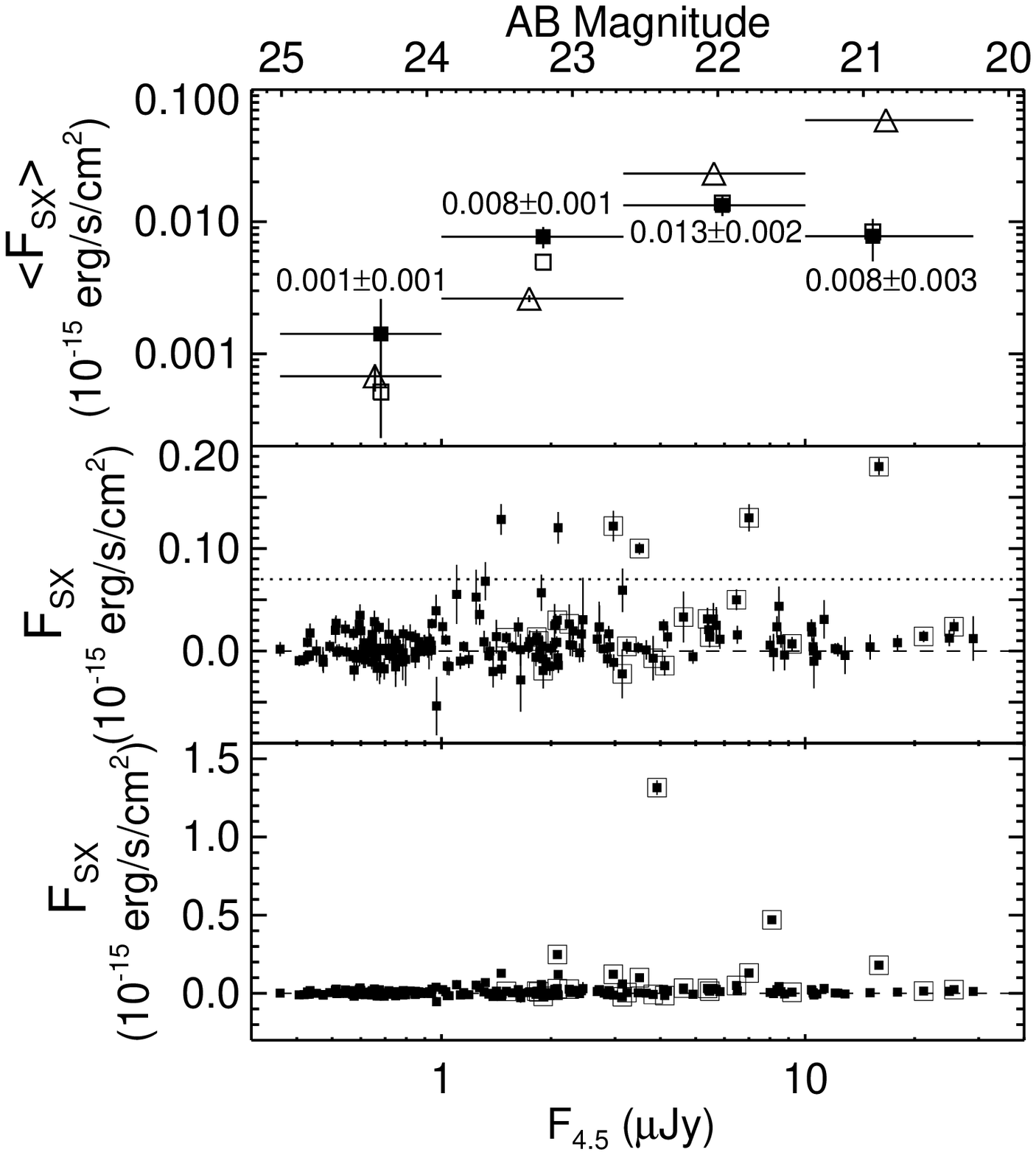}
\caption{Same as Figure~\ref{fig_hx} but in the soft X-ray (0.5--2.0 keV). 
The middle panel is a blowup of the bottom one.
\label{fig_sx}} 
\end{figure}

Unlike the radio image, the 2 Ms \emph{Chandra} observations do not probe into
the starburst regime at $z>2$, making it difficult to address whether the 
remaining 189 KIEROs contain significant X-ray AGN activity.
At $z=2$ and $z=3$, a $10^{42}$ erg s$^{-1}$ X-ray luminosity
corresponds to X-ray fluxes of $4.7\times10^{-17}$ and $1.7\times10^{-17}$
erg s$^{-1}$ cm$^{-2}$, respectively, for $\Gamma=1.8$.  The sensitivity of 
the 2~Ms observations do not reach such flux limits, especially in the hard X-ray
band (e.g., see error bars in Figures~\ref{fig_hx} and \ref{fig_sx}).  Thus, KIEROs undetected 
by \emph{Chandra} may still host $>10^{42}$ erg s$^{-1}$ X-ray AGNs.

%
%

%
%

We carried out stacking analyses on the hard (2--8 keV) and soft (0.5--2.0 keV) 
unfiltered X-ray images of \citet{alexander03} to understand the average X-ray properties 
of KIEROs.  We measured X-ray fluxes using aperture diameters approximately 
$2\times$ the PSF FWHM.  Since the FWHM of the \emph{Chandra} images vary 
with distance from the field center, we adjusted the size of the aperture 
accordingly.  We measured the background locally after masking the detected objects.
We compared our X-ray flux measurements with those given in the 2~Ms catalog and found
excellent agreement.  We estimated the errors (caused by confusion and shot noise) in the 
stacked fluxes with Monte Carlo simulations of random sources, as we did in the longer wavebands.  

The stacked hard and soft X-ray fluxes for all the 196 KIEROs are 
$\langle F_{HX}\rangle  = 2.9\pm0.9 \times 10^{-17}$
and $\langle F_{SX} \rangle = 8.1\pm1.4 \times 10^{-18}$ erg s$^{-1}$ cm$^{-2}$.
The hard X-ray value is close to the flux for X-ray AGNs, and the soft X-ray value falls 
significantly below.
Again, to understand if these stacked fluxes are dominated by small numbers of bright X-ray
sources, we applied flux cutoffs that are approximately $5\sigma$ in both hard and soft bands,
indicated by dotted lines in Figures~\ref{fig_hx} and \ref{fig_sx}. We imposed the same cutoff in the
Monte Carlo simulations, which allows us to greatly reduce errors caused by bright 
X-ray sources that get close to our targets by chance.
Furthermore, we stacked the X-ray fluxes of the KIEROs according
to their 4.5~$\mu$m fluxes, as we did in the longer wavebands.  
The results are presented in the top panels of Figures~\ref{fig_hx} and \ref{fig_sx}.

In Figure~\ref{fig_hx}, the stacked hard X-ray fluxes are consistent with AGNs
at $F_{4.5~\rm\mu m} > 3$ $\mu$Jy.  This suggests a significant AGN contribution
in the $F_{4.5~\rm\mu m}$ brightest members of KIEROs.  On the other hand,
KIEROs with $F_{4.5~\rm\mu m} < 3$ $\mu$Jy show no detectable hard X-ray fluxes,
and all the stacked soft X-ray fluxes in Figure~\ref{fig_sx} are below the flux required
for X-ray AGNs.  This suggest that the majority of KIEROs do not host X-ray AGNs.
In the following sections, we will further identify AGNs in the radio and mid-IR (MIR) and
exclude these AGNs in our analyses of star formation properties.

\subsection{MIPS 24~$\mu$m Properties}\label{sec_mips}

The \emph{Spitzer} GOODS Legacy Program DR1+ data release
provided a 24~$\mu$m catalog obtained with the \emph{Spitzer} Multiband 
Imaging Photometer \citep[MIPS;][]{rieke04}.  The catalog has a 24~$\mu$m 
flux limit of 80 $\mu$Jy and has 1041 sources in our IRAC region.
Among the 1041 sources, 18 are KIEROs.  The fraction of KIEROs in the
DR1+ 24~$\mu$m sample is thus 1.7\%, lower than the fraction in the radio (5.1\%
of the 1.4 GHz sample of \citealp{morrison10}, Section~\ref{sec_radio}).  
However, the 80 $\mu$Jy flux limit used in the DR1+ catalog is 
shallower than the true depth of the image.
To search for fainter MIPS detected KIEROs, we first made a deep source 
extraction in the MIPS image with SExtractor \citep{bertin96}.
We looked for matches between KIEROs and faint 24 $\mu$m sources
within $2\arcsec$ radii (approximately 1/3 of the PSF FWHM).  
We then inspected all the matches by eye to remove
unreliable matches caused by nearby bright sources and noise spikes.  
In particular, there are KIEROs residing in crowded regions with several
blended MIPS sources. We removed such KIEROs unless their are
clearly associated with local 24 $\mu$m peaks.
The above procedure increases the
number of MIPS detected KIEROs to 27. Simple aperture photometry indicates
that this pushes the 24 $\mu$m detection limit to slightly less than 50 $\mu$Jy,
corresponding to approximately 5 $\sigma$.  After this, the fraction of 24~$\mu$m 
detections among the sample of 196 KIEROs is 14\%, comparable to the fraction 
of radio detected KIEROs, which is 11\%. This similarity suggests that radio and 24 
$\mu$m observations are roughly equally effective in picking up KIEROs.

There is, however, an additional complication in the 24~$\mu$m case.
At $z>1$, multiple broad PAH emission and silicate
absorption features enter the MIPS 24~$\mu$m band, producing both 
positive and negative selection biases that are strong functions of redshift.
These strongly affect the observed 24~$\mu$m fluxes of KIEROs, since
they are all at $z>1$.  Because of this, using 24~$\mu$m fluxes to quantify the
star formation activity in KIEROs is highly uncertain \citep[see, e.g.,][]{murphy09,rodighiero10}
and strongly depends on the assumed SED models.  On the other hand, 
the SEDs of galaxies in the radio are simple power laws with very similar spectral indices, 
and the radio--FIR correlation is fairly insensitive to dust temperature.  We thus only relied 
on the radio data to study the SFRs of KIEROs.  

We did not attempt to probe deeper at 24 $\mu$m with stacking
analyses given the large uncertainties in interpreting the results.
Instead, we stacked the radio, millimeter, and submillimeter fluxes of the MIPS detected  and 
undetected KIEROs, to see whether the stacked fluxes are dominated by
the few 24 $\mu$m detected sources and hence whether there are also passive objects
in the KIERO selection.  First, the 27 MIPS detected KIEROs have mean 1.4 GHz, 
1100 $\mu$m, and 850 $\mu$m fluxes of $42.1\pm1.7$ $\mu$Jy (after excluding the $>600$ $\mu$Jy
bright radio source), $1.94\pm0.22$ mJy, and $6.15\pm0.68$ mJy, respectively.  
All these values are approximately $4\times$ higher than the mean fluxes of all KIEROs
listed in Table~\ref{tab_stack}.
This is not surprising since we expect the 24 $\mu$m detections to be very luminous at 
high redshift.  The remaining non-MIPS KIEROs have mean 1.4 GHz, 1100 $\mu$m, 
and 850 $\mu$m fluxes of $5.0\pm0.7$ $\mu$Jy, $0.27\pm0.09$ mJy, and
$0.45\pm0.31$ mJy, respectively.  The results are still significant, except at 850 $\mu$m
where the sample size is small (72).  Furthermore, if we exclude the
$F_{1.4~\rm GHz}>20~\mu$Jy sources from our radio stacking analyses, as we did in 
Section~\ref{sec_radio}, then the mean radio flux of the remaining 157 non-MIPS 
sources is $2.3\pm0.5$ $\mu$Jy.  This is comparable to the result 
of $2.6\pm0.5$ $\mu$Jy for all $F_{1.4~\rm GHz}<20~\mu$Jy sources.
We conclude that even sources undetected at 24 $\mu$m have
significant radio, millimeter, and submillimeter emission.  We do not find 
evidence for a passive population based on the 24 $\mu$m to 1.4 GHz data.

\section{AGN Fraction}\label{sec_agn}

In Section~\ref{sec_x}, we found seven KIEROs detected in the \emph{Chandra} 2 Ms 
catalog \citep{alexander03}.  All seven sources are X-ray AGNs with soft or hard X-ray luminosities greater than
$10^{42}$ erg s$^{-1}$.  To obtain a more complete identification of X-ray AGNs, we measured the
X-ray fluxes of KIEROs as we did in the X-ray stacking analyses. We found four additional sources with 
$>3\sigma$ X-ray fluxes and with X-ray luminosities exceeding $10^{42}$ erg s$^{-1}$.
Therefore, there are 11 X-ray AGNs in our KIERO sample.

In the MIR, our previous work \citep{barger08,wang10} had shown that the IRAC AGN selection
techniques of \citet{lacy04} and \citet{stern05} will either be incomplete or highly contaminated
by normal galaxies in a deep sample like this. Here we adopt the MIR power-law selection technique
\citep{donley07}. We require that the objects have to be detected in all four IRAC bands, and
we performed power-law fitting to their IRAC SEDs.  Among the 196 KIEROs, 54 are bright enough
to be detected in all IRAC bands, and 13 have SEDs that can be well fitted 
($\chi^2$ probability $>90\%$) by red power laws.  
All of them have red IRAC spectral slopes with $\alpha > 0.7$, consistent 
with them being AGNs.  
One of the 13 is detected by \emph{Chandra} and is also an X-ray AGN.

In the radio, we estimated the rest-frame 1.4 GHz power of radio detected KIEROs using their 
photometric redshifts and assuming a synchrotron spectral slope of $\alpha=0.8$.  We found two KIEROs with unusually large 1.4~GHz radio powers of $>10^{25}$ W Hz$^{-1}$.
Neither is a MIR power-law AGN.
One source is \#58 in Figure~\ref{fig_sed}, an X-ray AGN discussed in Section~\ref{sec_x}.
Its photometric redshift is very likely to be a catastrophic failure caused by its unusual SED, but
this does not affect its AGN identification. The other source is \#70 in Figure~\ref{fig_sed}.  
Its photometric redshift fitting appears reasonable, and thus the estimated rest-frame radio 
power is reliable.  Its large radio power ($2.1\times10^{25}$ W Hz$^{-1}$) requires a SFR of 
$>10^4~M_\sun ~{\rm yr}^{-1}$ (see next section), which is not supported by its
infrared properties (e.g., it is undetected at 24 $\mu$m).  We thus conclude that its 
radio emission is powered by an AGN.

The above X-ray, MIR, and radio selection identify 23 AGNs, which is 12\% of the KIEROs in the
GOODS-N. This fraction is a lower limit, since AGNs with lower X-ray, MIR, or radio luminosities
would not be selected. On the other hand, all 23 AGNs are detected in the IRAC bands.
Thus, the AGN fraction of the 54 members of the
IRAC-bright subsample is $\sim40\%$.  This sets an upper limit on the
AGN fraction of the entire KIERO sample since we expect to have more AGNs in luminous sources.
\citet{messias10} studied AGNs in high-redshift extremely red galaxy populations using radio, X-ray, and
MIR criteria.  They found an AGN fraction of up to $\sim30\%$, among which the majority are
obscured MIR power-law AGNs.  Their results are consistent with ours, given the small sample size here.

To see if the above AGN identification picks up X-ray AGNs that are below the \emph{Chandra} 
detection limit, we repeat the hard X-ray stacking analyses in Figure~\ref{fig_hx} without all the
above AGNs.  In the 4.5 $\mu$m flux ranges of $>10$ $\mu$Jy and 3--10 $\mu$Jy, 
the stacked hard X-ray fluxes decrease substantially to $4.0\pm2.5\times10^{-17}$ and
$0.3\pm2.1\times10^{-17}$ erg s$^{-1}$ cm$^{-2}$, respectively.  Both values are not statistically
significant.  Therefore the above X-ray, MIR, and radio AGN identification is fairly effective in removing potential 
X-ray AGNs.


\begin{deluxetable*}{lccccc}[ht!]
\tablewidth{0pt}
\tablecaption{Star Formation Properties of KIEROs \label{tab_sfr}}
\tablehead{\colhead{Redshift} & $N_{\rm galaxy}$ & \colhead{$\langle F_{1.4~\rm GHz} \rangle$} & 
\colhead{$\langle L_{IR} \rangle$} & \colhead{$\rm \langle SFR \rangle$} & \colhead{SFRD} \\ 
& & \colhead{($\mu$Jy)} & \colhead{($10^{12}L_\sun$)} & \colhead{($M_\sun$ yr$^{-1}$)} & \colhead{($M_\sun$ yr$^{-1}$ Mpc$^{-3}$)} }
\startdata
\multicolumn{6}{c}{All Sources} \\
\hline
$z\sim1.5$--2 	& 10		& $51.0\pm2.7$	& 3.4		& 580	& 0.016 	\\
$z\sim2$--3 	& 27 		& $13.4\pm1.7$   	& 2.2		& 370 	& 0.013	\\
$z\sim3$--4 	& 23		& $18.5\pm1.8$	& 6.7		& 1150 	& 0.036	\\
$z\sim4$--5.5 	& 10 		& $31.3\pm2.7$  	& 20.8	& 3532 	& 0.053	\\
unknown		& 120	& $2.5\pm0.8$		& \nodata	& \nodata & \nodata \\
\hline
\multicolumn{6}{c}{Non-AGNs} \\
\hline
$z\sim1.5$--2 	& 9		& $46.5\pm2.9$	& 3.1		& 530 	& 0.013	\\
$z\sim2$--3 	& 22 		& $14.5\pm1.9$   	& 2.4		& 400 	& 0.011	\\
$z\sim3$--4 	& 15		& $18.9\pm2.2$	& 6.9		& 1170 	& 0.024 	\\
$z\sim4$--5.5 	& 7 		& $7.3\pm3.3$  	& 4.8		& 820 	& 0.009	\\
unknown		& 115	& $1.9\pm0.8$		& \nodata	& \nodata	& \nodata 
\enddata
\end{deluxetable*}

\section{Star Formation Rate}\label{sec_sfr}

The SFRs of the KIEROs can be inferred from their total IR luminosities.
Among all the flux measurements in Table~\ref{tab_stack}, the radio fluxes provide
the most robust estimates of the FIR luminosities. This is because the radio spectral slopes of normal 
galaxies are very similar and there is a tight correlation between radio and IR luminosity 
of normal galaxies, regardless of dust temperature (see the review of \citealp{condon92}).  
The latest FIR observations suggest that this correlation also holds for high-redshift galaxies 
\citep[e.g.,][]{ivison10a,ivison10b,bourne11}.  An additional reason for using the
radio is the higher S/N in the stacked radio fluxes.  

A possible uncertainty of using radio fluxes for SFR estimates is the contribution from AGNs.
In the previous section, we identified AGNs using X-ray, MIR, and radio data.
This allows us to exclude AGNs from our sample.  Furthermore, most of the KIEROs have radio 
fluxes well below 100 $\mu$Jy.  At $z\sim3$ this flux corresponds to a radio
power of $6\times10^{24}$ W Hz$^{-1}$.  In the local universe, AGNs dominate the radio luminosity function at
$P_{1.4~\rm GHz}>10^{24}$ W Hz$^{-1}$ \citep[e.g.,][]{mauch07}.  On the other hand, at $z>1$,
starbursts start to dominate sources around $10^{24}$ W Hz$^{-1}$, and the luminosity function of AGNs
falls substantially below that of starbursting galaxies at lower radio powers \citep[e.g.,][]{cowie04}.
Thus, it is plausible that the contributions from AGNs to the radio fluxes of our KIERO sample  are negligible.
Below we will show this is indeed the case using AGNs identified in the previous section.

We now derive the conversion between radio flux and SFR.
The radio power can be calculated from
\begin{equation}
P_{1.4~\rm GHz}=4 \pi d_L^2(1+z)^{\alpha-1}F_{1.4~\rm GHz},
\end{equation}
where $d_L$ is the luminosity distance and $\alpha$ is the radio spectral slope.
Here we assume $\alpha=0.8$.  To calibrate the conversion between radio power and SFR,
we assume the radio--FIR correlation
\citep{condon92}
\begin{equation}
q \equiv \log\left( \frac{L_{\rm FIR}}{3.75\times10^{12} \rm W}\right)
- \log\left( \frac{P_{1.4~\rm GHz}}{\rm W Hz^{-1}}\right)
\end{equation}
and the canonical $q$ value of 2.3.  Using the SED templates in \citet{silva98}, we found that 
for a broad range of FIR SEDs the total IR luminosity conventionally defined between 8 and 1000~$\mu$m is 
approximately $2\times$ that of the 40--120~$\mu$m FIR luminosity.  
With this conversion factor and the standard conversion
between total IR luminosity and SFR in \citet{kennicut98},
\begin{equation}
{\rm SFR} ~(M_\sun ~{\rm yr}^{-1}) =1.7\times10^{-10} L_{\rm IR} (L_\sun), 
\end{equation}
we derived
\begin{equation}
{\rm SFR} ~(M_\sun ~{\rm yr}^{-1}) = 7.83\times10^{-8} (1+z)^{\alpha-1} \left(\frac{d_L}{Mpc}\right)^2 \left(\frac{F_{1.4~\rm GHz}}{\mu \rm Jy}\right).
\end{equation}
This agrees with the conversion in \citet{yun01} to within 10\% and is $\sim20\%$ 
higher than that in \citet{bell03}.

In Table~\ref{tab_sfr} we present the star formation properties of the KIEROs 
based on the stacked radio fluxes.
The upper half of the table is for all KIEROs, and the lower half is for KIEROs 
without X-ray, MIR, and radio AGNs.
It can be seen that excluding AGNs only slightly decreases the mean radio fluxes, 
except in the highest redshift
bin, where the stacked radio fluxes are significantly biased by the two radio AGNs. 
This supports the argument that
most of the KIEROs have radio emission powered by star formation.  
Below we will limit our discussion to the non-AGN results.

For the sources with redshifts, the SFRs and IR luminosities are similar to 
those of local ultraluminous IR galaxies 
\citep[ULIRGs;][]{sanders96} such as Arp 220. They are also close to the 
lower end of typical dusty galaxies 
found by millimeter and submillimeter single-dish surveys 
($L_{\rm IR}\gtrsim10^{13}L_\sun$). This is what we expected
for the KIERO color selection.  It is difficult to estimate the SFRs of 
the KIEROs without redshifts.
If their redshift distribution is similar to that of KIEROs with redshifts, 
then they would be roughly an order of magnitude
less luminous, or luminous IR galaxies (LIRGs).  
On the other hand, it is possible that these $K_S$ and
IRAC faint sources are at higher redshifts.  
A mean radio flux of 1.9 $\mu$Jy from $z\sim3$--4 would imply
a SFR of $\sim10^3M_\sun~{\rm yr}^{-1}$ and a ULIRG luminosity.  
We can thus safely conclude that
most of the KIEROs are at least LIRGs at high redshifts, 
and many of them are starbursting ULIRGs.

We investigated whether there is a correlation between stellar mass and SFR.
We did this only in the $z=2$--4 range because of the sample size.  For $z=2$--3, we divided the non-AGN KIEROs
into three groups according to their stellar masses: $<10^{10}M_\sun$ (7 sources), $10^{10}$--$10^{11}M_\sun$ (11 sources), and 
$>10^{11}M_\sun$ (7 sources).  Their stacked radio fluxes are, respectively, $12.7\pm3.3$, $5.6\pm2.6$, and $23.3\pm3.3$ $\mu$Jy.
For $z=3$--4, we divided them into two groups: $<10^{11}M_\sun$ (6 sources), and $>10^{11}M_\sun$ (9 sources).
Their stacked radio fluxes are, respectively, $19.2\pm2.9$ and $18.4\pm3.5$ $\mu$Jy.
There is not a convincing trend here, except for fluctuations likely caused by the small sample sizes.
Therefore, even the most massive KIEROs in our sample
are still actively forming stars, and there is no evidence for a significant number of
massive and passive galaxies in the KIERO population.  This is quite different than other
high-redshift red galaxies. We believe this is because our KIERO color selection avoids the
4000~\AA~Balmer  break (a signature of old stars) and primarily picks up galaxies with 
continuous red spectral slopes (a signature of dust extinction).

\begin{figure}[h!]
\epsscale{1.0}
\plotone{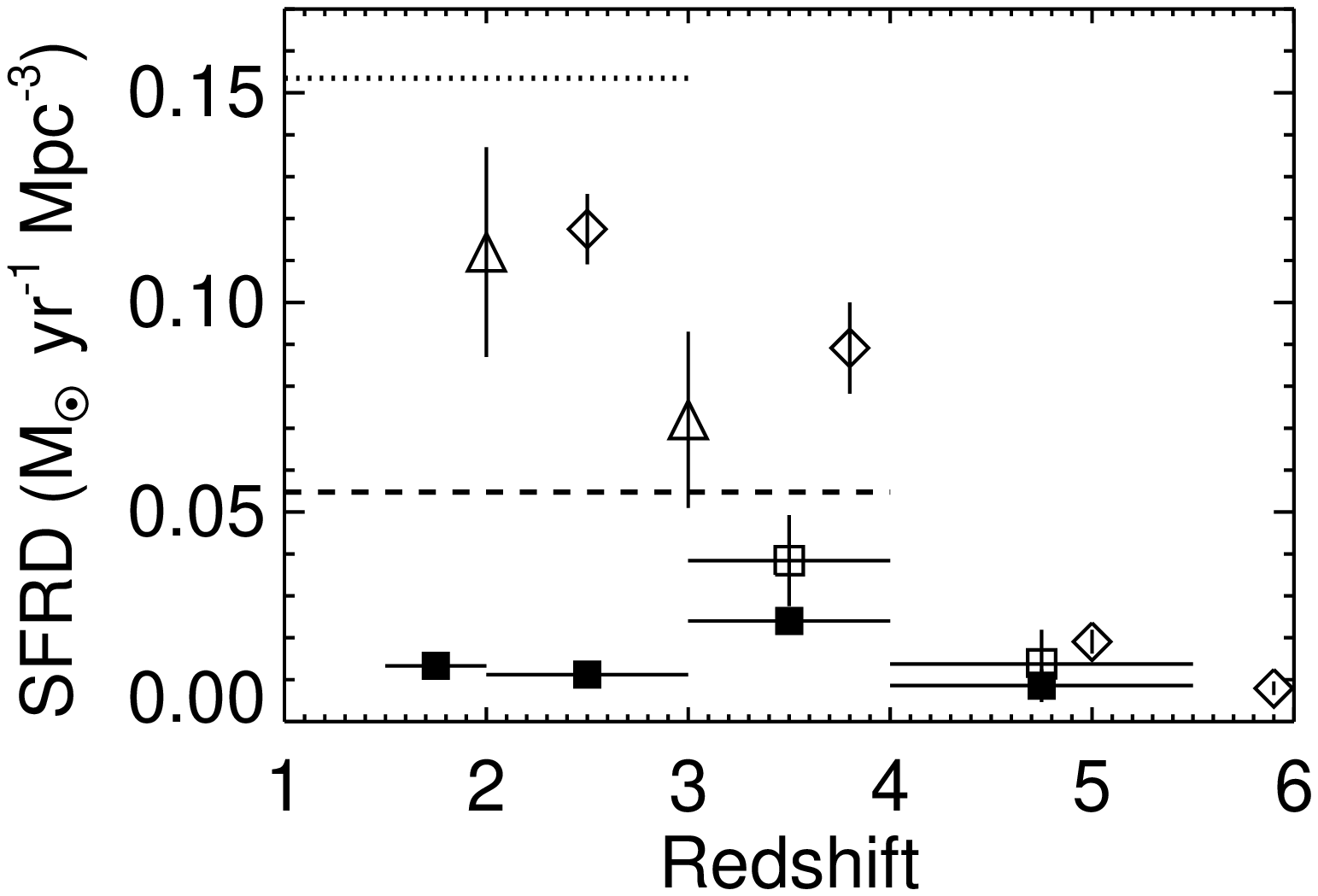}
\caption{Comoving SFRD.  Solid squares are SFRDs derived from 
our radio stacking analysis on KIEROs with redshift information.
Open squares further include the contributions from KIEROs without redshifts,
by assuming $z>3$ for these $K_S$ and IRAC faint KIEROs (see text for details).
Open diamonds and triangles are extinction corrected rest-frame UV 
results from \citet{bouwens09} and \citet{reddy09}, respectively, for LBGs integrated
to 0.04 $L^\star$.  The dashed line shows the mean SFRD in Wang et al.\ (2006) at $z=1$--4, 
derived using a radio stacking analysis on a large NIR selected sample.  It is approximately flat.
The dotted line shows the SFRD in Wang et al.\ (2006) at $z=1$--3 assuming the maximum possible
completeness correction in the submillimeter.
\label{fig_sfrd}} 
\end{figure}

We estimated the contributions of KIEROs to the cosmic star formation, i.e., their SFRD.
Normally, one would like to apply the $1/V_{max}$ method \citep{schmidt68} for calculating volume.
Here, since most of our sources are not individually detected in the radio, applying such a method
is effectively introducing additional weighting according to their $K_S$ or 4.5 $\mu$m properties 
in the radio stacking analyses.  It is unclear to us whether this would bias the results.
Therefore, we simply adopt total comoving 
volumes in the four redshift bins in Table~\ref{tab_sfr} for our survey
area (0.06 deg$^2$). The derived SFRD for the sources with redshifts are listed in the table
and shown in Figure~\ref{fig_sfrd} (solid squares).
These values are lower limits for the following reasons:  (1) they are not corrected for 
the substantial incompleteness in the $K_S$ and 4.5~$\mu$m bands; (2) they do not
include star formation in the KIEROs identified as AGNs; and (3) they do not include contributions 
from the KIEROs without redshifts.  
If we assume that the redshift distribution of the KIEROs without redshifts
is similar to that of the KIEROs with redshifts, then the SFRD values in Table~\ref{tab_sfr} 
would increase by 20\% in all four redshift bins. 
On the other hand, if we assume that the KIEROs without redshifts
are, on average, from higher redshifts, then their contribution to 
the SFRD would be substantial.
For example, if we assume that they are at $z>3$, then the SFRDs at $z>3$ 
would increase by 60\% (open squares in Figure~\ref{fig_sfrd}).

The SFRD values we found are comparable to those for bright SMGs
at similar redshifts \citep{chapman05}, but we note that not all the bright 
SMGs are KIEROs.
In Wang et al.\ (2006), we found a nearly flat SFRD of $\sim0.06$--0.15 
$M_\sun$ yr$^{-1}$ Mpc$^{-3}$ at $z=1$--4, from our radio and 850~$\mu$m stacking analyses
of a large 1.6 and 3.6~$\mu$m selected NIR sample (dashed line in Figure~\ref{fig_sfrd}, while
the dotted line shows the same SFRD maximally corrected for incompleteness in the $z=1$--3 range). 
Naively speaking, the KIEROs
should be a subsample of the larger NIR sample of Wang et al.\ (2006).
However, the CLEAN-like procedure employed in W10 for this work provides many more
3.6 and 4.5~$\mu$m detected sources and better photometry.
By comparing the list of KIEROs with our 3.6 $\mu$m catalog compiled in 2006, 
we found that 45\% of the KIEROs in the SCUBA area were missed by Wang et al.\ (2006).
The missed KIEROs have a broad range of 3.6 $\mu$m fluxes and the same IRAC image is
used in both works, so this is not an effect of different detection limits.  
Because of this, comparing the SFRD here with that in Wang et al.\ (2006) is not straightforward.

%
%

We also compare our SFRDs with recent extinction corrected rest-frame UV 
determinations from LBGs (\citealp{bouwens09}, open diamonds in Figure~\ref{fig_sfrd};
\citealp{reddy09}, open triangles).  At $z>3$, the SFRD of the 
KIEROs is similar in magnitude to the SFRD of the extinction corrected LBGs.
At $z<3$, the KIERO SFRD contribution becomes only $\sim10\%$ of 
that of LBGs.  This might be explained by the fact that KIEROs do not include all SMGs
and therefore they only account for part of the dusty star formation. However, even the 
radio/submillimeter SFRD determined by stacking on a large NIR sample by 
Wang et al.\ (2006; dashed line in Figure~\ref{fig_sfrd}) 
is significantly lower than the SFRD of the extinction corrected LBGs.  

The extinction corrected SFRD of LBGs seems to be unusually large at $z\gtrsim2$, 
compared to submillimeter and KIERO SFRDs.  In Table~\ref{tab_stack} we 
showed that optically faint KIEROs are brighter in the radio, millimeter, and submillimeter than KIEROs 
detected by ACS. In Section~\ref{sec_lbg}, we find that KIEROs have very little overlap with LBGs. 
One important finding in Wang et al.\ (2006) is that most of the submillimeter EBL arises from galaxies 
with intermediate rest-frame colors, i.e., galaxies not as blue as LBGs.  
The combination of these points suggests that a substantial amount of high-redshift star formation
cannot be traced by rest-frame UV emission, and that the total SFRD should be the sum of 
the SFRD from LBGs and the SFRD from dusty sources.  This would give a very large total SFRD at 
$z\gtrsim2$, which raises the question of whether the 
large extinction corrected LBG SFRD is overestimated.

\citet{carilli08} stacked radio fluxes of $z\sim3$ LBGs in the COSMOS field and found that the
radio inferred mean SFR of these LBGs is only 1.8 times the \emph{extinction uncorrected} mean UV SFR.
They pointed out that this is much smaller than the factor of $\sim5$ generally adopted 
for high-redshift LBGs.  Furthermore, \citet{ho10} could not detect the radio signal
from $z\sim4$ LBGs with an extremely deep radio stacking analysis in the GOODS-N and S fields.
Their inferred radio SFR of $z\sim4$ LBGs is only consistent with the extinction corrected UV SFR in
\citet{bouwens09} if a 2~$\sigma$ radio flux upper limit is adopted.  This result also suggests a discrepancy
between the radio SFR and the dust-corrected UV SFR.  The reason for this discrepancy is unclear. 
In addition to the possibilities discussed in \citet{carilli08}, another possibility is the correlation between 
FIR flux and UV spectral slope \citep{meurer99} assumed in many high-redshift LBG studies.
Recent observations show substantial scatter in this correlation \citep{cortese06,howell10,wijesinghe11}, and this may
contribute to the uncertainty in the UV extinction correction.  This issue will require 
further investigation, especially once direct measurements of the FIR luminosities of 
LBGs with sensitive instruments such as ALMA and \emph{Herschel} become available.

\section{KIERO and Other High-Redshift Populations}\label{sec_overlap}

\subsection{IRAC Selected EROs}\label{sec_iero}

\citet{yan04} used the ACS $z^{\prime}$ (F850LP) image and the IRAC 3.6~$\mu$m 
image in the HUDF and found 17 IRAC selected 
extremely red objects (IEROs) with $z^{\prime}-3.6~\mu\rm m > 3.25$ \citep[also see][]{yan08}.  
Among our 196 KIEROs, 123 have $z^{\prime}$ detections at $>3\sigma$, and only 27 of 
them are above the IERO color cut.  If we adopt the $3\sigma$ limits for $z^{\prime}$ undetected 
KIEROs, then the number of IEROs increases to 68.  Thus, $>1/3$ of KIEROs are also IEROs.
Among the 17 IEROs in Yan et al., five satisfy 
the KIERO color cut, including one $K_S$ undetected IERO. 
Thus, approximately 1/3 of IEROs are KIEROs.
The cumulative density of IEROs with $K_S<25$ is $\sim1.4$ arcmin$^{-2}$,
more than $2\times$ higher than that of KIEROs even after a $K_S$ based completeness correction.
From these points of view, KIEROs and IEROs are two different populations with roughly
30\% overlap.
  
On the other hand, the KIEROs are remarkably similar to IEROs in many other properties.
First, the 17 IEROs in Yan et al.\ have redshifts between 1.6 and 3.6, 11 of them
between 2 and 3.  This is almost identical to the redshift distribution for KIEROs
(Figure~\ref{fig_photoz}a).  Secondly, the SEDs of IEROs mostly consist of 
strong NIR components from massive populations of old stars, plus 
secondary rest-frame UV components from ongoing star formation.
This is also very similar to the SEDs of KIEROs (Figure~\ref{fig_sed}
and Section~\ref{sec_redshift}), except that some KIEROs are too faint in the optical
for the rest-frame UV components to be detected. The stellar masses derived
from the SED fitting for IEROs and KIEROs are also similar, approximately
$10^{9.5}$--$10^{11.5} M_\sun$.  In Section~\ref{sec_data}, we showed that 14 out of 
the 28 identified SMGs in the GOODS-N are KIEROs.  If we include $J$-band upper limits,
then 22 out of the 28 SMGs are IEROs. This fraction is even higher than KIEROs.

Furthermore,  \citet{messias10} performed radio stacking analyses on several red galaxy populations
including IEROs. They found a mean radio flux of $8.2\pm1.9$ from 96 IEROs at $z=2$--3 after
excluding 3 sources detected at $>3\sigma$, where 1 $\sigma$ is $\sim14$--17 $\mu$Jy
in their observations.  With a similar $\sim50$ $\mu$Jy cut in our stacking of 
$z=2$--3 KIEROs, the stacked radio flux is $7.5\pm1.4$ $\mu$Jy, quite comparable to
IEROs at similar redshifts. The only difference here is that nearly 10\% of KIEROs 
(2 out of 27) are above the 50~$\mu$Jy flux cut, but only 3\% of IEROs in \citet{messias10} 
are above this cut, indicating that the KIERO population contains more ULIRGs.
However, within the numerical uncertainties, these two percentages could be consistent.

\begin{figure}[h!]
\epsscale{1.0}
\plotone{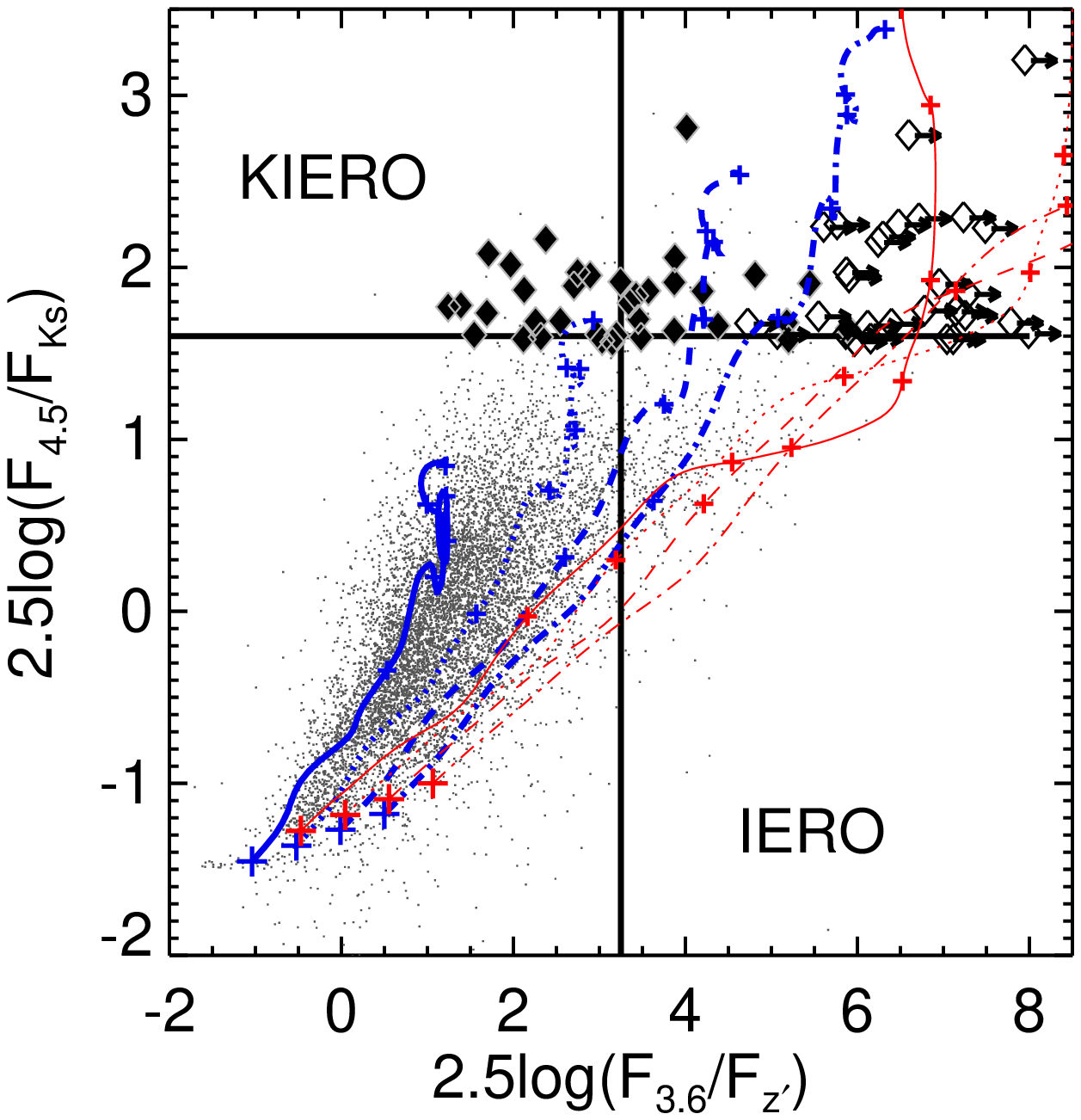}
\caption{$z^{\prime}$, $K_S$, and IRAC color--color diagram.  The vertical and horizontal lines
indicate the color criteria for IEROs and KIEROs, respectively.  Color curves show the 
tracks of elliptical galaxies (thin red curves, Coleman et al.\ 1980) and starburst galaxies 
\citep[thick blue curves,][]{kinney96},
reddened with the \citet{calzetti00} extinction law with $A_V=0$ (solid), 1.0 (dotted), 2.0 (dashed),
and 3.0 (dash-dotted).  Integer redshifts are indicated by crosses along the tracks, starting from
$z=0$ in the lower-left and ending at $z=6$.  Solid diamonds show KIEROs detected at both $K_S$
and $z^{\prime}$. Open diamonds with arrows show 1~$\sigma$ $z^{\prime}-3.6$ $\mu$m limits of 
$z^{\prime}$-undetected KIEROs.
Dots in the background are field galaxies that 
are detected at $z$, $K_S$, and 3.6~$\mu$m at $>5$~$\sigma$.
\label{fig_iero}} 
\end{figure}

Why two such similar populations do not overlap much in terms of color properties
and number densities is an interesting question.  In the color--color diagram in 
Figure~\ref{fig_iero} we see that the IERO selection and the KIERO selection overlap
at high redshift on dusty sources.  The difference is in the low-redshift end. 
In the lower-right part of the diagram, early-type galaxies (including those not forming stars) 
enter the IERO selection at $z>1.5$ or so.  Such galaxies do not enter the KIERO selection 
at $z<3$.  The upper-left part of the diagram is more interesting, i.e., galaxies that are
KIEROs but not IEROs.  Such galaxies have blue $z^{\prime}-3.6~\mu\rm m$ colors that
are consistent with starbursting galaxies at $z>6$, simply based on the color tracks.
These KIEROs are not $z>6$ galaxies.  Instead, as described in Section~\ref{sec_stellar},
they have two distinct stellar populations.  Their massive and reddened stellar populations 
make them redder than the KIERO criterion, and the unobscured parts of their ongoing starbursts 
make them bluer than the IERO criterion.  In short, IEROs contain more early-type galaxies,
and KIEROs contain more dusty sources with ongoing star formation.  We believe this is 
a consequence of the fact that the KIERO selection avoids the Balmer break but targets the
red spectral slopes caused by strong dust extinction.

%
%

\subsection{Distant Red Galaxies}\label{sec_drg}

In addition to IEROs, distant red galaxies \citep[DRGs, $J-K_S>1.35$;][]{franx03} 
are another interesting population with which to compare.  
DRGs have redshifts roughly between 2 and 4 \citep[e.g.,][]{vandokkum03,forster04,reddy05}, 
similar to KIEROs and IEROs.  We therefore expect a substantial overlap between 
the DRG and KIERO populations.  Unfortunately, there is not yet a $J$-band image
in the GOODS-N that is sufficiently deep and wide to verify this.  The CFHT WIRCam $J$-band
image that we mentioned in Section~\ref{sec_redshift} covers the entire GOODS-N with an rms depth of 
0.09 $\mu$Jy (cf.\ 0.12 $\mu$Jy at $K_S$).  Of the 104 $K_S$ detected KIEROs, only 37 are detected 
in this $J$-band image, and of those, only six have $J-K_S$ colors redder than 1.35.  
An additional 15 $J$ undetected KIEROs have $J$-band flux upper limits ($3\sigma$) satisfying the DRG color.
Thus, at least 52 of the 104 $K_S$ detected KIEROs are DRGs, and the DRG fraction
among the $K_S$ detected KIEROs can be anywhere between 50\% and 70\%.  

There exists another 
$J$-band image around the HDF-N taken with the 8.2-m Subaru Telescope \citep{kajisawa06}.
It is substantially deeper than the CFHT image.  We carried out an independent reduction of this 
image \citep{wang07}, which covers $\sim26$ arcmin$^2$ with an rms depth of $\sim0.03$ $\mu$Jy.
There are 16 $K_S$ detected KIEROs in this image.  Five of them are detected in this deep 
$J$ image but all of them do not satisfy the DRG criterion.  Five of the remaining six $J$-band
undetected sources have $J$ upper limits that satisfy the DRG criterion.  Thus, the fraction of
$K_S$ detected KIEROs that are DRGs is anywhere between 30\% and 70\% based on this 
narrower and deeper $J$ image.

Finally, in the extremely deep \emph{HST} NICMOS F110W image of the HDF-N 
(e.g., \citealp{dickinson00}; rms $\sim0.15$ $\mu$Jy), 
there are only three KIEROs.  Only one of them is detected in the $K_S$-band and it is not
a DRG. However, this image is too narrow for us to draw a meaningful conclusion.

\begin{figure}[h!]
\epsscale{1.0}
\plotone{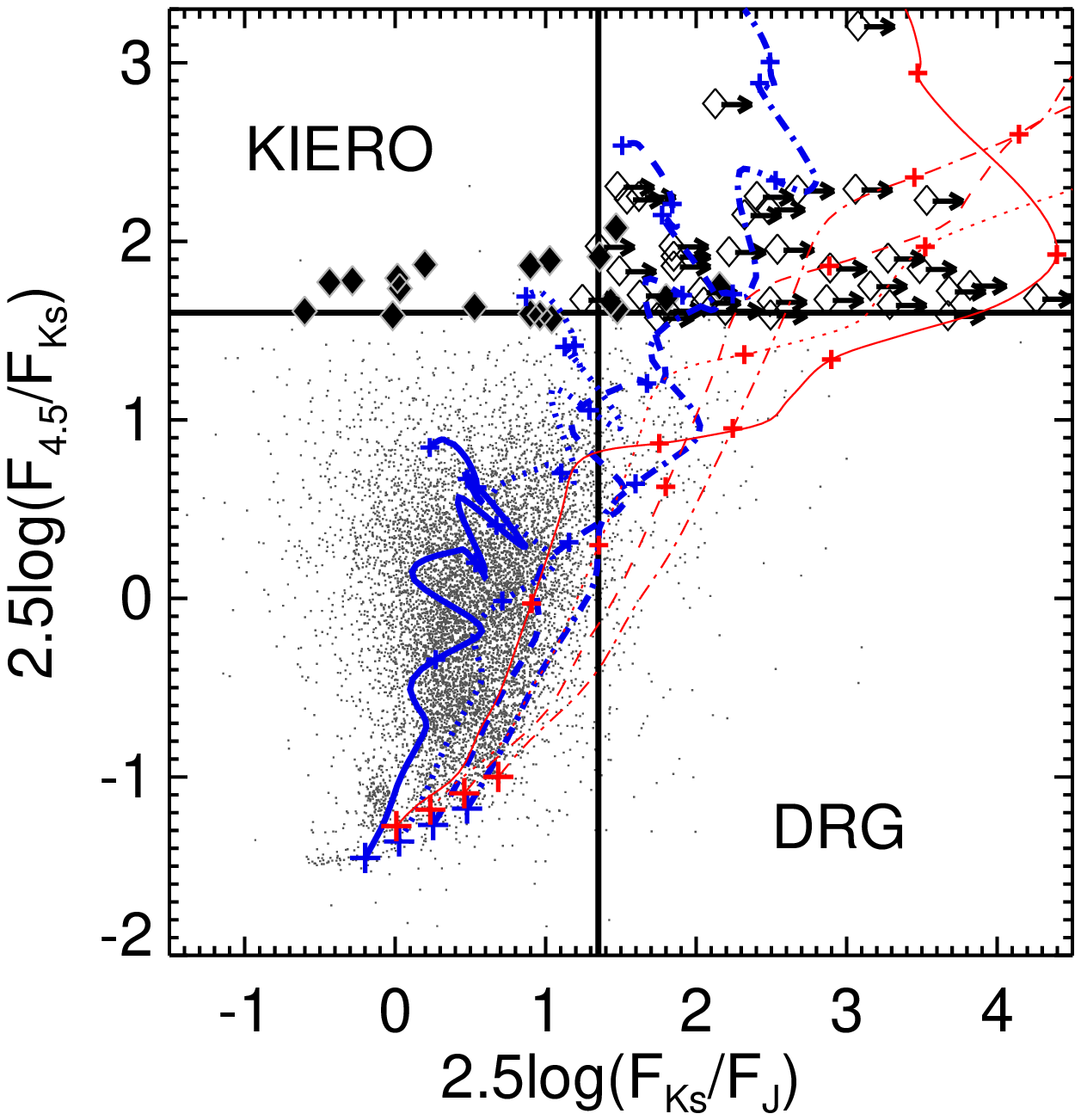}
\caption{$J$, $K_S$, and 4.5 $\mu$m color--color diagram.  The vertical and horizontal lines
indicate the color criteria for DRGs and KIEROs, respectively.  Curves are the same as those in
Figure~\ref{fig_iero}.  Solid diamonds show KIEROs detected in both $K_S$
and $J$-band WIRCam images. Open diamonds with arrows show 1~$\sigma$ $J-K_S$ limits of 
$J$-undetected KIEROs.  Dots in the background are field galaxies that are detected at
$J$, $K_S$, and 4.5 $\mu$m at $>5$~$\sigma$.
\label{fig_drg}} 
\end{figure}

In terms of number density, the
cumulative density of DRGs at $K_S<24$ is $\gtrsim2$ arcmin$^{-2}$ 
\citep{labbe03,grazian06b,kajisawa06}, significantly higher than that for KIEROs
(0.9 arcmin$^2$ at our $K_S$ and 4.5~$\mu$m limits).
However, there is tentative evidence that the number density of DRGs starts to drop at 
$K_S>24$ \citep{kajisawa06}, while that of KIEROs still seems to be
increasing at this magnitude (Section~\ref{sec_density}).  If both observed trends are
real, KIEROs are not just the faint-end tail of the DRGs.  
Based on the above results, we conclude that most DRGs are not KIEROs.
On the other hand, a substantial fraction (perhaps $>50\%$) of KIEROs are DRGs,
but not all.  In Figure~\ref{fig_drg} we show the $J$, $K_S$, and 4.5 $\mu$m color--color
diagram.  It shows that the difference between DRGs and KIEROs is somewhat 
similar to that between IEROs and KIEROs.

%
%

In the SMG population, the fraction of DRGs is high.  Among the 28 identified SMGs
in the GOODS-N, 15 are DRGs.  This fraction is comparable to the KIERO fraction in SMGs.
\citet{knudsen05} studied the SFRs of DRGs in a lensing cluster field
with a submillimeter stacking analysis at 850~$\mu$m.  For DRGs with $K<24.4$ they found a 
mean 850~$\mu$m flux of 0.93 mJy and a mean SFR of 127 $M_\sun$ yr$^{-1}$ (both 
corrected for lensing amplification).  Based on this, the typical SFRs of KIEROs
are $\gtrsim2\times$ higher than those of DRGs, despite being fainter in the NIR.
On the radio side, \citet{messias10} performed radio stacking analyses on DRGs.
They found a mean radio flux of $6.1\pm1.6$ $\mu$Jy from 152 DRGs at $z=2$--3 after 
excluding 3 sources detected at $3\sigma$. As mentioned in Section~\ref{sec_iero},
a similar stacking of $z=2$--3 KIEROs gives a comparable 7.5 $\mu$Jy radio flux,
but the KIERO population has a higher ULIRG fraction.

\subsection{Lyman Break Galaxies}\label{sec_lbg}

To find LBGs in the 112 members of the ACS-detected subsample of KIEROs,
we adopted the selection criteria of \citet{beckwith06} for galaxies in the HUDF.
The selection criteria are also similar to those used by \citet{bouwens07}.  
We found 15 $b$-dropouts, 4 $v$-dropouts, and 5 $i$-dropouts.  
This shows that most of the KIEROs are not LBGs, even for the subsample that
is detected in the optical.
The reason for this is that the spectral slopes are too red to be selected by the
dropout criteria.  Furthermore, the number of LBGs ($\sim1700$ $b$-dropouts in GOODS-N; 
\citealp{bouwens07}) is much higher than the number of KIEROs (197 in GOODS-N).  
All these results imply that KIEROs and LBGs are two distinct populations.  

It is important to realize that KIEROs are the most intensive starbursting galaxies at $z>2$, 
second only to the brightest SMGs.  Their high SFRs (Section~\ref{sec_sfr}) imply strong
rest-frame UV radiation.  Without absorption, their redshifted optical fluxes would be roughly 
1.6--2 $\mu$Jy at 0.5--1~$\mu$m.  However, even in the ACS-detected KIERO subsample, 
most of them have optical fluxes lower than this (e.g., Fig.~\ref{fig_sed}).  
The factor of almost 10 in the rest-frame UV dust attenuation in 
KIEROs is much larger than the generally accepted extinction corrections of 
$\sim3\times$ to $5\times$ in LBGs \citep[e.g.,][]{bouwens07,reddy08,bouwens09}.

\subsection{BzK Galaxies}\label{sec_bzk}

\begin{figure}[h!]
\epsscale{1.0}
\plotone{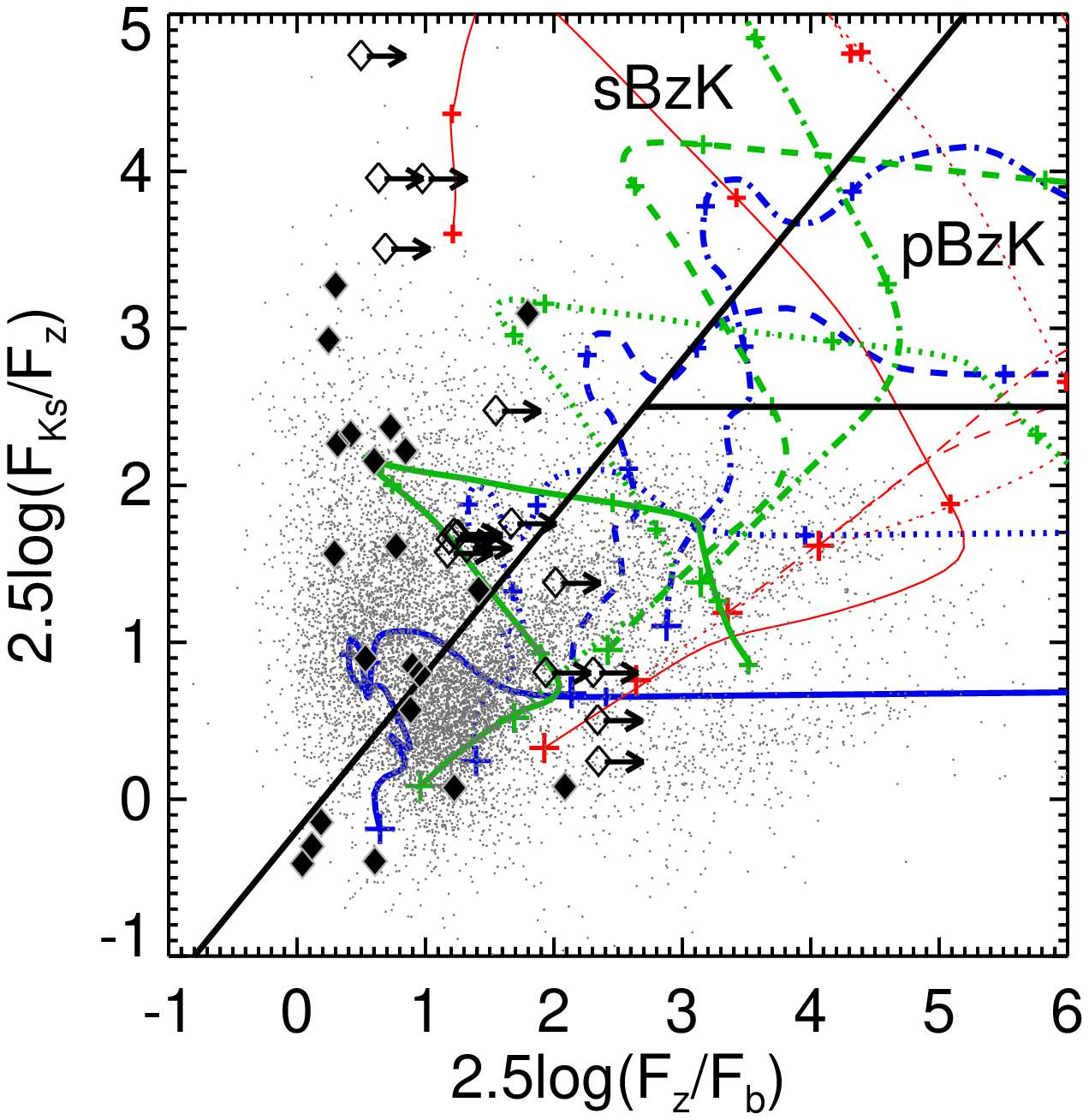}
\caption{$b$, $z^\prime$, and $K_S$ color--color diagram.  
The black diagonal and horizontal lines
are the selection criteria for sBzK and pBzK galaxies.  
Curves are the same as those in
Figure~\ref{fig_iero}, except that we added Scd-type galaxies (green curves, Coleman et al.\ 1980).  
Solid diamonds are KIEROs detected at $K_S$, $z^{\prime}$, and $b$ bands.
Open diamonds with arrows show 1~$\sigma$ $z^{\prime}-b$ limits of $b$-undetected KIEROs.
Dots in the background are field galaxies that are detected in all three of the
$b$, $z^\prime$, and $K_S$ bands at $>5$~$\sigma$.  
Squares are KIEROs that were detected in all three bands at $>3$~$\sigma$.
Triangles show the lower limits for the KIEROs that are undetected in the $b$-band.
\label{fig_bzk}} 
\end{figure}

The BzK selection \citep{daddi04} is an effective technique for finding 
$1.4\lesssim z \lesssim 2.5$ galaxies.  Since this redshift range overlaps with 
the lower-end of the redshift distribution of KIEROs,
it is interesting to see where KIEROs fall on the BzK diagram (Figure~\ref{fig_bzk}). 
A difficulty here is that most KIEROs are optically faint, and we can only put a small 
subsample of KIEROs on
the BzK diagram.  Only 21 KIEROs are detected in all three of the $b$, $z^\prime$, and 
$K_S$ bands (squares), 14 of which are starburst BzK galaxies (sBzK). 
Sixteen KIEROs are detected in both the $z^\prime$ and $K_S$ bands (triangles), 
11 of which have $b-z^\prime$ lower limits consistent 
with sBzK.  The non-BzK KIEROs either have
photometric redshifts that are too high or too low for the BzK selection (\#15, 35, 53, and 55 in 
Figure~\ref{fig_sed}) or have unusually blue rest-frame UV and optical SEDs (\#44 and 63 in 
Figure~\ref{fig_sed}) that barely miss the BzK selection.  

We conclude that roughly 2/3 of the optically bright KIEROs are sBzK galaxies.
However, the requirement of $b$ and $z^\prime$-band data makes the BzK selection
less effective for extremely optically faint populations like the KIEROs.
Finally, among the 28 identified SMGs, 13 are sBzK galaxies.  This fraction is 
similar to the KIERO fraction in SMGs.  However, as shown by \citet{pannella09} and \citet{kurczynski11},
the BzK selection picks up much more ``normal'' star forming galaxies with SFR 
$\sim10$--100 $M_\sun$/yr.
The SMG fraction in BzK galaxies is much lower than the SMG fraction in KIEROs.

%
%

\section{Discussion}\label{sec_discuss}

\subsection{Contributions to the Extragalactic Background Light}

Our original goal of selecting extremely red sources from the $K_S$ and IRAC 4.6~$\mu$m
bands was to see if we could select high-redshift dusty sources in the NIR.  
In this paper (Section~\ref{sec_sfr}) we showed that KIEROs are mostly LIRGs 
and ULIRGs at $z>2$.  We found that KIEROs are optically faint.  Their red optical 
SEDs imply that they will not be picked out as LBGs, even 
if they are detected in the optical.  All of the results indicate that we have successfully 
selected a population of galaxies that are too dusty to be selected by existing optical 
and NIR surveys and too faint to be systematically 
detected with existing MIR, millimeter, and submillimeter surveys.

An interesting question is whether  the KIERO sample includes the majority of ULIRGs 
at $z>2$.  At 850~$\mu$m and 1100~$\mu$m, KIEROs have an integrated surface brightness of 
$4.47\pm0.88$ and $1.64\pm0.26$ Jy deg$^{-2}$ (Table~\ref{tab_stack}), respectively.  
The EBL measured by the \emph{COBE} satellite has strengths
of 31--44 Jy deg$^{-2}$ at 850~$\mu$m and 18--25 Jy deg$^{-2}$ at 1100~$\mu$m. 
These were measured by two groups \citep{puget96,fixsen98}, and the range reflects the uncertainties
in removing foreground contamination.
Thus, KIEROs contribute approximately 10\% to the EBL at these wavelengths 
and do not seem to represent the majority of sources that 
give rise to the millimeter and submillimeter EBL.  

On the other hand, there is evidence that the majority of the
millimeter and submillimeter EBL (especially at $\ge 850$~$\mu$m) 
arises at $z<2$ (Wang et al.\ 2006; \citealp{serjeant08,marsden09}),
and thus that the KIERO selection may still pick up a large fraction 
of $z>2$ dusty sources.  For example, Wang et al.\ (2006) found an 
850~$\mu$m EBL contribution of $\sim16$ Jy deg$^{-2}$ from
a sample of 1.6 and 3.6~$\mu$m selected sources at $z<2$.  This leaves at most
15--28 Jy deg$^{-2}$ to sources at $z>2$, of which $\sim15\%$--27\% can be
attributed to the KIERO sample here.  The actual fraction should be even larger 
considering the fact that the KIERO selection is still severely limited by completeness 
at $K_S$ and 4.5~$\mu$m.
To determine whether or not KIEROs can fully account for the EBL that arises at $z>2$, we will
need both a better determination of the KIERO number counts and a better EBL measurement.

\subsection{Dust Hidden Star Formation}

In Section~\ref{sec_lbg} we showed that the dust attenuation of the rest-frame UV radiation 
of KIEROs is roughly a factor of 10 or even higher.  It is now known that some of the most luminous 
millimeter or submillimeter selected galaxies can be entirely hidden by dust in the optical and even in the 
NIR \citep[e.g.,][]{wang09,cowie09}.
The work presented here shows that there are still extremely extinguished
galaxies when we  go roughly an order of magnitude fainter in the total IR luminosity.
Our KIERO color selection picks up such galaxies at $z\sim1.5$--5,
but the $K_S$ and IRAC images become less sensitive to galaxies at $z>3$.  With a narrow selection
window between $z\sim2$ and 4 and the completeness limits in the $K_S$ and the 4.5 $\mu$m bands, 
we already pick up roughly 10\% of the total EBL at
850~$\mu$m and 1100~$\mu$m.  The total fraction of background that arises from extremely 
extinguished galaxies from $z=0$ to $>6$ should be even higher than this.  
We thus expect a significant fraction
($\gg10\%$) of cosmic star formation to be entirely hidden from deep optical observations.

\subsection{Nature of KIEROs}
The KIERO color selection is meant to pick up objects with red spectral slopes over broad
wavelength ranges and to avoid the 4000~\AA~Balmer  break.
This makes it sensitive to dusty starbursting galaxies.
The fact that the KIEROs are selected near the peak of their stellar SEDs (i.e., the rest-frame
1.6~$\mu$m bump) also means that it is sensitive to massive systems.  
In Section~\ref{sec_redshift} we showed that most KIEROs 
have stellar masses of $10^{10}$ to $10^{12}$ $M_\sun$.  
In Section~\ref{sec_sfr} we showed that most KIEROs are ULIRGs with $L_{\rm IR}>10^{12}L_\sun$
and are actively forming stars. The masses of KIEROs are very similar to those of bright 850~$\mu$m 
selected SMGs at $z>2$ (e.g., \citealp{dye08}).  
On the other hand, the mean 850~$\mu$m flux
of KIEROs (1.44 mJy) is below the confusion limit of single-disk telescopes. 
Its corresponding IR luminosity (assuming a dust SED similar to Arp 220)
are a few times lower than those of most SMGs selected at
850~$\mu$m by single-dish telescopes. This suggests that 
KIEROs are massive galaxies undergoing 
slightly less intensive starbursts compared to bright SMGs.  

In Section~\ref{sec_data} we found that half of the identified SMGs in the GOODS-N are KIEROs.
The fraction of SMGs that are DRGs is comparable (Sections~\ref{sec_drg}), and 
the fraction of SMGs that are IEROs is significantly higher ($\sim80\%$, Section~\ref{sec_iero}).  
However, this does not mean that all of these color selections are equally efficient in picking up 
dusty sources.  The IERO and DRG selections pick up
substantial numbers of passive and $z<2$ galaxies \citep[see also,][]{messias10}.
However, in our radio stacking analyses of KIEROs, we did not see evidence for a significant
subpopulation of passive KIEROs.  This is true even after we introduce a 24 $\mu$m selection
to our radio stacking analyses (Section~\ref{sec_mips}).
It is especially remarkable that approximately 1/3 of KIEROs
are not detected by ACS.  Such UV-faint sources would be considered as ``passive'' by
many studies, but these KIEROs are even brighter in the radio than UV-bright KIEROs, indicating
that they are more active in star formation.  This distinguishes the KIERO selection from the 
IERO and DRG selections. We believe the KIERO selection is more effective in picking up
dusty starburst galaxies and contains fewer passive galaxies.

In this paper we provide estimates of stellar masses and SFRs of KIEROs. 
Another crucial parameter for understanding the evolution is gas mass, which is not yet
established for this new population of galaxies.  Given the great similarity (SFRs, dust obscuration,
and stellar masses) between KIEROs and bright SMGs, it is likely that the two 
populations are similarly massive in molecular gas mass.  Recent observations 
of high-$J$ CO transitions of SMGs found molecular gas masses of a few $10^{10} M_\sun$
\citep{neri03,greve05,tacconi06,tacconi08}.  If KIEROs are similarly rich in molecular gas, 
they can sustain their ultraluminous starburst phase for $\sim100$ Myr.  In other words, 
they can significantly increase their stellar masses in less than $10\%$ of their Hubble time.

To summarize, the KIERO selection seems to pick up a sample of massive and dusty $z>2$ galaxies. 
They are forming stars actively. This is consistent with massive galaxies at $z>2$ that are still rapidly building up their mass. The
properties of KIEROs are similar to their more luminous counterparts (SMGs), as well as to high-redshift
massive galaxies selected in the IRAC bands (e.g., IEROs and DRGs; see also \citealp{mancini09}).
In each of these various selections we are likely seeing different regions of the
luminosity function or different stages in the formation of the most massive galaxies at $z>2$.  
Eventually these 
galaxies will become quiescent and passively evolve into the ``downsizing'' era.  However, at
high redshifts, most of them seem to be still active.  Because of the dust obscuration, common for
the most intensive star-forming systems, a complete view of such massive and active galaxies
requires observations at both long and short wavelengths.

\section{Summary}\label{summary}

In order to find high-redshift, faint, dusty sources that may give rise to a significant fraction of the
submillimeter EBL, we selected 196 KIEROs with extremely red colors of $K_S-4.5$~$\mu$m$>1.6$ in the 0.06 deg$^2$ 
GOODS-N region.  The selected KIEROs have a range of 4.5~$\mu$m fluxes, mostly between 1 and 10
$\mu$Jy. Of the 196 KIEROs, 104 KIEROs are detected in the $K_S$ band, and 
the slope of the $K_S$ number counts for these sources is steeper than
that of $K_S$ selected galaxies at similar flux levels, likely because KIEROs are at redshifts higher
than $K_S$ selected galaxies.  The counts also still seem to be rising at the $K_S$
detection limit of $\sim0.3$ $\mu$Jy.  Roughly 2/3s of KIEROs are detected in the optical by \emph{HST}
ACS.  However, only 1/5 of the ACS-detected KIEROs are LBGs.  The remaining ACS
detected KIEROs all have 
optical continua that are too red to pass the LBG selection.

We performed a photometric redshift analysis on a $K_S$-bright subsample of 76 KIEROs. We found a 
redshift distribution between $z\sim1.5$--5, with roughly 70\% at $z\sim2$--4.  This redshift distribution
is quite similar to that of bright SMGs, but with a bias to high redshift.  Our photometric redshift analysis
also shows large stellar masses of $10^{10}$--$10^{12} M_\sun$.  We found that a significant fraction of
KIEROs have SEDs that cannot be well fitted by simple stellar populations.  They have massive old
stellar components that are luminous in the NIR and ongoing starbursts that are detected in the
optical. 

We found that roughly half of the identified millimeter and submillimeter sources in the GOODS-N 
are selected as KIEROs, but such sources only represent 7\% of the selected KIEROs.  To study the 
radio, millimeter, and submillimeter properties of the remaining KIEROs, we carried out stacking analyses.  
We found that, on average, KIEROs are brighter than normal 4.5~$\mu$m selected galaxies by factors of 
$\sim2$ and $\sim6$--10 in the radio and in the millimeter and submillimeter, respectively.  We also found that
KIEROs undetected by ACS are $\sim3$--4 times brighter in all these wavebands than KIEROs detected by ACS.
We repeated the stacking analyses on KIEROs undetected in the radio, millimeter, and submillimeter.
These faint KIEROs still possess detectable stacking signal, showing that their properties are not 
dominated by small numbers of bright sources.  We also did not find evidence of passive KIEROs based
on the  24 $\mu$m to radio data.

We identified AGNs in the KIERO sample with large X-ray and radio luminosities, and with MIR
SEDs that are featureless red power laws.  We identified 19 such sources, of which 2/3 are obscured
AGNs identified in the MIR.  We also found an AGN fraction of up to 1/3.

Using the local radio--FIR correlation of normal galaxies and the stacked radio fluxes of non-AGN KIEROs, 
we found that they are LIRGs and ULIRGs.  The results are consistent with the KIEROs being a 
population of $z>2$ ultraluminous dusty galaxies that are slightly 
fainter than SMGs in the millimeter and submillimeter and undetected by single-dish surveys.
Their large IR luminosities imply high SFRs of $\sim300$--1200 $M_\sun$ yr$^{-1}$,
and therefore a population of massive $z>2$ galaxies in rapid formation.
We showed that KIEROs contribute roughly 10\% to the total EBL at 850 and 1100~$\mu$m
and up to 30\% of the EBL at these wavelengths which arises from galaxies at $z>2$.  
Given that KIEROs and LBGs have very little overlap and that
optically faint KIEROs are brighter in the radio, millimeter and
submillimeter, these results indicate that a significant fraction 
($\gg10\%$) of cosmic star formation is hidden by dust and 
cannot be traced by rest-frame UV emission.

\acknowledgments
We thank the referee for the thorough review and very useful comments.
We gratefully acknowledge support from the National Science
Council of Taiwan grants 98-2112-M-001-003-MY2 and
99-2112-M-001-012-MY3 (W.H.W.), NSF grants AST 0708793 (A.J.B.) 
and AST 0709356 (L.L.C.),
the University of Wisconsin Research Committee with funds
granted by theWisconsin Alumni Research Foundation
and the David and Lucile Packard Foundation (A.J.B.).

\end{document}